\documentclass[fleqn,usenatbib]{mnras}
\usepackage{graphicx}
\usepackage{hhline}
\usepackage[mathscr]{euscript}
\usepackage{outlines}
\usepackage{xcolor}
\usepackage{placeins}
\usepackage{amssymb, amsmath}
\newcommand{\rns}{\rho_{\rm sat}}
\newcommand{\Ms}{M_{\odot}}
\newcommand{\beq}{$\beta$-equilibrium}

\newcommand{\ROP}{R\"{O}P~}
\usepackage{tikz}
\usepackage{pgfplots}
\usepackage{hyperref}

\title[]{Finite-temperature effects in dynamical spacetime binary neutron star merger simulations:\\ Validation of the parametric approach}

\author[C. Raithel, P. Espino, \& V. Paschalidis]{
Carolyn A. Raithel,$^{1,2,3}$\thanks{E-mail: craithel@ias.edu}
Pedro Espino,$^{4,5}$
Vasileios Paschalidis,$^{6,7}$
\\
$^1$School of Natural Sciences, Institute for Advanced Study, Princeton, NJ 08540, USA\\
$^2$Princeton Center for Theoretical Science, Princeton University, Princeton, NJ 08540, USA\\
$^3$Princeton Gravity Initiative, Princeton University, Princeton, NJ 08540, USA \\
$^4$University of California Berkeley Department of Physics, Berkeley, CA 94720, USA \\
$^5$Institute for Gravitation and the Cosmos, The Pennsylvania State University, University Park, PA 16802, USA\\
$^6$Department of Astronomy and Steward Observatory, University of Arizona, Tucson, Arizona 85721, USA \\
$^7$Department of Physics, University of Arizona, Tucson,  Arizona 85721, USA 
}

\date{Accepted XXX. Received YYY; in original form ZZZ}

\pubyear{2022}

\begin{document}
\label{firstpage}
\pagerange{\pageref{firstpage}--\pageref{lastpage}}
\maketitle

\begin{abstract}
Parametric equations of state (EoSs) provide an important tool for systematically studying EoS effects
in neutron star merger simulations. In this work, we perform a numerical validation of the $M^*$-framework for
parametrically calculating finite-temperature EoS tables. The framework, introduced in Raithel et al. (2019),
provides a model for generically extending any cold, \beq~EoS to finite-temperatures and arbitrary
electron fractions. In this work, we perform numerical evolutions
of a binary neutron star merger with the SFHo finite-temperature EoS, as well as with the 
$M^*$-approximation of this same EoS, where the approximation uses the zero-temperature, \beq~
slice of SFHo and replaces the finite-temperature and composition-dependent
 parts with the $M^*$-model. We find that the approximate version of the EoS
is able to accurately recreate the temperature and thermal pressure profiles of the binary neutron 
star remnant, when compared to the results found using the full version of SFHo.
We additionally find that the merger dynamics and gravitational wave
signals agree well between both cases,
with differences of $\lesssim1-2\%$ introduced into the post-merger gravitational wave peak frequencies
by the approximations of the EoS. 
We conclude the $M^*$-framework can be reliably used to probe neutron star merger
properties in numerical simulations.
\end{abstract}


\maketitle

\section{Introduction}

Binary neutron star mergers provide a promising new avenue for studying the 
dense-matter equation of state (EoS) across a wide range of conditions. 
During the early inspiral, the neutron stars are thermodynamically cold
and the interior matter remains in \beq.
As the neutron stars come into contact with one another,
shock heating raises the temperature of the system to $\mathcal{O}(10)$s of MeV \citep[e.g.,][for
  reviews]{Baiotti2017,Paschalidis2017}, at which point the thermal pressure is significant and
  can influence the evolution of the post-merger remnant
 \citep[][]{Oechslin2007,Baiotti2008,Bauswein2010,Bauswein2010a,Sekiguchi2011,Paschalidis2012,Raithel2021}. 
 At these temperatures, the matter can also deviate significantly from equilibrium
\cite[e.g.,][]{Rosswog:2003rv,Sekiguchi2011,Hammond:2021vtv,Most2021}, and out-of-equilibrium effects may
 become important for some EoSs \cite[e.g.,][]{Most2022}.
If the remnant object avoids prompt collapse to a black hole, the massive 
neutron star remnant will additionally probe matter at extreme densities and masses not accessible by isolated neutron stars. 
Such conditions provide a unique laboratory for studying
the dense-matter EoS.

When it comes to exploring realistic EoS effects in neutron star merger simulations,
there are two main approaches. The first is to use a
tabulated, finite-temperature EoS, which can be calculated with a variety of methods,
ranging from the liquid-drop model of \citet{Lattimer1991}, 
to the relativistic mean-field (RMF) approach of \citet{Shen1998a}.
Another ten models have been computed with the statistical model of 
\citet{Hempel2012} for different RMF models and nuclear mass tables,
 while many more finite-temperature EoSs are being added to the available libraries, 
thanks in part to the CompOSE database which enables public sharing of such
 tables \citep{Typel2015}. These microphysical EoS tables provide a robust 
 method for testing the predictions of a
 particular theory with a given set of nuclear parameters, 
 coupling constants, and calculation methods. For a review of
finite-temperature EoSs, see \citet{Oertel2017}. 

However, for general comparisons of neutron star merger properties, 
there are some drawbacks to limiting our studies to the existing sample of tables.
For example, the existing catalog of finite-temperature EoS tables currently
includes few models that predict neutron stars with radii $\lesssim 12$~km,
whereas recent constraints from low-mass X-ray binary
observations and from GW170817
provide significant evidence for more compact stars,
with radii between $\sim11$ and 13~km.
\citep[for reviews, see][]{Ozel2016,Baiotti2019,Raithel2019a,Chatziioannou2020}.

Additionally, when comparing simulation results that 
use existing finite-temperature EoS tables, there are multiple differences between
the tables that complicate straightforward comparisons. For example, these
tables vary not only in their predictions for the neutron star compactness or maximum
mass, but they also differ in the thermal pressure that they predict at
a given temperature \citep[see Fig. 1 of][]{Raithel2021}. While it is well understood
that differences in the cold physics (affecting, e.g., the stellar compactness) can
influence the post-merger evolution, differences in the finite-temperature
 part of the EoS can also influence post-merger properties such as the ejecta and
 the gravitational wave emission \citep{Bauswein2010,Figura2020,Raithel2021}. 
When using pre-existing EoS tables in simulations, it can be difficult 
to disentangle the effects of changing these properties
simultaneously. 

In order to get around these limitations, a second approach has been 
developed to study EoS effects in neutron star mergers more systematically. 
In this approach, a cold EoS is extended to finite-temperatures with an approximate 
prescription, which can be held fixed or varied, independently of the cold EoS. 
The cold EoS could be a microphysical EoS tabulated at zero-temperature, 
of which there are many more options 
than in the finite-temperature case \citep[e.g.,][]{Ozel2016}; or it could be an agnostic
parametrization, such as piecewise polytropes \citep{Read2009,Ozel2009}, 
which can be
designed to probe a new part of the parameter space. 
In early merger simulations, it was common to approximate the
thermal extension of the EoS with a constant thermal
index, according to $P_{\rm th} =\epsilon_{\rm th} (\Gamma_{\rm th}-1) $, where $P_{\rm th}$ 
and $\epsilon_{\rm th}$ are the thermal pressure and energy density, respectively, 
and $\Gamma_{\rm th}$ is a constant \citep{Janka1993}. 
For $\Gamma_{\rm th}=5/3$, this so-called ``hybrid approach" is equivalent to an ideal-fluid 
prescription. For more realistic EoSs, the thermal index is expected to vary
 significantly with the density, as the matter becomes degenerate \citep{Constantinou2015a}.

In \citet{Raithel2019,ROPerratum} (hereafter \ROP), a new framework was developed for 
extending cold, \beq~EoSs to finite-temperatures and arbitrary electron fractions. 
In that work, the high-density thermal prescription is based on a two-parameter approximation
 of the particle effective mass, in order to account for the effects
of degeneracy on the thermal pressure at high-densities, and thus to 
provide a more realistic density-dependence for the effective thermal index. The framework of
 \ROP also allows for the initial EoS to be extended from \beq~
to arbitrary electron fractions, using a parametrization of the nuclear symmetry energy. 
In \ROP,  it was shown that for a sample of nine published, finite-temperature EoS tables based on
relativistic energy density functionals, 
the $M^*$-framework was able to re-create the pressure of the complete models
with errors of $\lesssim$30\%, at densities and temperatures of interest for neutron star mergers.
The $M^*$-framework reduces the error of the thermal
pressure model compared to the ideal-fluid based approximation of the hybrid approach by up to 3-4 orders of magnitude. The first successful numerical implementation of the $M^*$-framework and application to binary neutron star mergers was presented in~\cite{Raithel2021} (see also~\citealt{Raithel:2022san} where the implementation was added to different cold EoS parametrization frameworks).

In this paper, we provide a complementary validation of the $M^*$-framework, by demonstrating that
the numerical simulation of a binary neutron
star merger evolved with an $M^*$-approximated EoS can recreate the results
that are found with the full version of the EoS.
 We do so using the finite-temperature EoS table SFHo \citep{Steiner2013}. 
 In particular, we take the zero-temperature,
 \beq~ slice of SFHo, and extend it to finite-temperatures and arbitrary electron fractions
 using the $M^*$-framework; thereby replacing the finite-temperature and composition-dependent
 parts with the $M^*$-model  at high densities. We perform evolutions of binary neutron star mergers
 with the full version of SFHo and with its $M^*$ approximation, and we confirm that the
 $M^*$-approximation accurately recreates the results found with the full EoS.

We note that this comparison is purely a validation of the framework, rather than its intended use.
That is, if one's goal is to use a particular, existing dense-matter model (such as SFHo), then
the existing published table can and should be used. The advantage of the $M^*$-model is that it 
allows for new EoSs to be constructed in new parts of the parameter space.
In this paper, we approximate SFHo with the simple goal of validating that the $M^*$-framework
is able to recreate realistic merger evolutions. We focus in particular
on diagnostics and observable properties that are sensitive to the 
high-density EoS,
which is the regime the $M^*$-framework is designed to approximate
(in contrast to the hybrid approach, which breaks down at high densities).
To that end, we confirm that evolutions with the $M^*$-framework lead to realistic
thermal profiles of the merger remnant, and that they reproduce
the post-merger dynamics
and gravitational wave emission
 predicted by an existing, tabulated EoS. 
  
 The outline of the paper is as follows. In Sec.~\ref{sec:EoS}, we summarize
 the construction of the approximate EoS table. In Sec.~\ref{sec:methods}, we describe
  the numerical set-up for our simulations. We present the simulation results
 in Sec.~\ref{sec:results}. We discuss the implications of these findings and conclude 
 in Sec.~\ref{sec:conclusion}. Unless otherwise indicated, we use natural 
 units in which $G=c=k_B=1$.

\section{Parametric modeling of the finite-temperature EoS}
\label{sec:EoS}
We start with a short overview of the construction of the approximate EoS 
table.
 We do not repeat all the details of the parametric model for the thermal 
 and composition-correction terms. The complete framework can be found 
 in \cite{Raithel2019,ROPerratum}. Here, we focus in particular on the details of the construction
 that are most relevant for the comparison to an existing, tabulated EoS.
 
  \subsection{The SFHo EoS}
We validate the $M^*$-framework against the SFHo EoS \citep{Steiner2013}, 
which was calculated within the statistical framework of \cite{Hempel2010} 
with a new set of relativistic mean field parameters designed to match 
neutron star observations. In particular, SFHo predicts the 
radius of a 1.4~$\Ms$ neutron star to be 11.89~km and the maximum mass to be 2.06~$\Ms$,
making it among the softest finite-temperature EoSs that are currently available.
 
In addition to its compatibility with observational data, we also choose SFHo as it
poses a stringent ``stress test" for modeling finite-temperature effects in a merger. This is because stars that 
are very compact, like those predicted by SFHo, reach shorter separations before 
merging, leading to more violent collisions and, accordingly, a higher degree of shock
 heating \citep[e.g.,][]{Bauswein2013}. Additionally, for a soft EoS such as SFHo,
 because the overall cold pressure is relatively low, any
thermal pressure contributes a greater fraction of the total pressure 
 than would be the case for a stiffer EoS.
  Thus, by validating the $M^*$-framework
 with SFHo, we are testing a challenging part of the parameter
space, where errors in the thermal framework should have the strongest impact 
on observable features.  This allows us to place an upper bound on how the 
approximations of the $M^*$-framework might influence
observable features, such as the gravitational wave signal.

 \subsection{Construction of 3D, approximate EoS tables}

\begin{figure}
\centering
\includegraphics[width=0.45\textwidth]{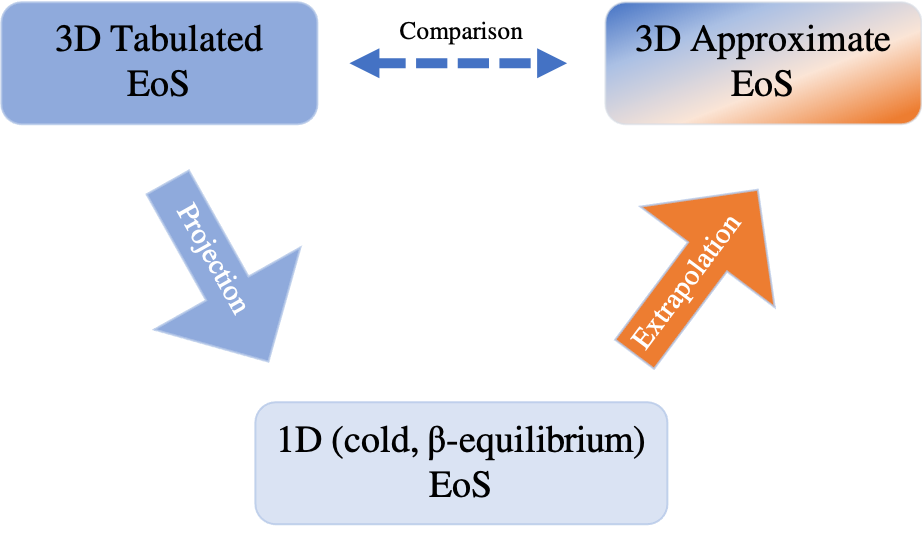}
\caption{\label{fig:construction} Cartoon schematic of the construction of the approximate 3D tables.}
\end{figure}

With this goal in mind, we construct the following test, which is summarized 
in Fig.~\ref{fig:construction}. 
We start by extracting a 1D constant-temperature slice  from the full table for SFHo, which we obtain from
 \texttt{stellarcollapse.org}. We label this process ``Projection" in Fig.~\ref{fig:construction}
 and we choose the 1D slice to correspond to the EoS
at approximately zero-temperature\footnote{In practice, the lowest temperature reliably included in 
the published tables is $T=0.1$~MeV. At this temperature, the thermal pressure
is completely subdominant compared to the cold pressure, rendering thermal effects negligible. Thus,
$T=0.1$~MeV is a reasonable approximation of ``cold" matter.} and in \beq. 
 We then extend the 1D slice of the EoS to finite-temperatures and arbitrary electron fraction
 using the complete $M^*$-framework.\footnote{We note that, throughout this work, we use 
 the term the ``$M^*$-framework" to refer to the complete model of \ROP, including both the finite-temperature
 \textit{and} composition-dependent correction terms outlined in eq.~(\ref{eq:Poutline}).}
 In skeletal format, this extension is performed according to 
 \begin{equation}
 \label{eq:Poutline}
 P(n, T, Y_e) = P(n, T\approx 0, Y_{e}^{\beta}) + P_{\rm th}(n, T, Y_e) + P_{\rm sym}(n, Y_e)
 \end{equation}
 where $P$ is the pressure, $n$ is the number density,
 $T$ is the temperature, $Y_e$ is the electron fraction, 
 and $Y_{e}^{\beta}$ is a short-hand notation to indicate the electron fraction for cold matter
 in \beq~(we note that $Y_{e}^{\beta}$ is in fact density-dependent, but we suppress this
 dependence for clarity). Thus, in Eq.~(\ref{eq:Poutline}), the first term is simply the
 cold, \beq~EoS; the second term represents the thermal correction term;
 and the third term represents the composition-correction 
 term, which extends the matter to non-equilibrium compositions.

 The \ROP model for the thermal pressure, $P_{\rm th}$, depends on
 two free parameters, $n_0$ and $\alpha$, 
 which govern the density-dependence of the particle effective mass function, $M^*(n)$. 
  Here,
 we use the \ROP fit coefficients for SFHo, $n_0=$0.22~fm$^{-3}$ and $\alpha=$0.89.
 \footnote{These values correspond to the best-fit parameters characterizing $M^*(n)$
  for symmetric nuclear matter ($Y_e=0.5$), fit together at $T=1,10,47.9$~MeV. 
 The effective mass function for SFHo was provided via the website of
  M. Hempel (priv. communication).
 See \ROP for details.} 
   
 The composition-correction term is constructed from a parametrization
 of the nuclear symmetry energy. The model for $P_{\rm sym}$ depends
 on the standard symmetry energy expansion 
 parameters $S_0$ and $L_0$, which are related to the magnitude and slope of the 
 symmetry energy at the nuclear saturation density; as well as a free parameter, $\gamma$, 
 which extrapolates the symmetry energy model to higher densities.
In constructing the approximate version of SFHo, we use the published SFHo
 values of $S_0$=31.57~MeV and  $L_0$=47.10~MeV \citep{Steiner2013},
 and the fit value of $\gamma=$0.41 provided in \ROP.
 For the complete expressions for $P_{\rm th}(n, T, Y_e)$ and 
 $P_{\rm sym}(n, Y_e)$, as well as the corresponding expressions for the energy 
 and the sound speed, see \cite{Raithel2019}.
 
Finally, we note that the $M^*$-framework is designed to
approximate thermal and composition-dependent effects in the high-density regime 
(i.e., above the nuclear saturation density, $n_{\rm sat}=0.16$~fm$^{-3}$). 
We do not attempt to use the $M^*$- model
at very low densities where (1) the physics is well understood and there is no need for
approximate models to explore new parts of the parameter space, and (2) the symmetry energy
approximation is complicated by the emergence of nuclei, for which the assumption
of homogenous nuclear matter no longer applies. For further discussion of the latter point and
for the solution we adopt (which involves applying a power-law tail, to ensure the symmetry energy 
correction term goes smoothly to zero at low densities), see Appendix A in \citet{Most2021}.
For these reasons, we switch to the full version of the EoS table in the low density
regime. The transition between the low-density EoS table and our high-density, approximate table
is performed using the free-energy matching procedure of \cite{Schneider2017}, which ensures
that the merged EoS remains thermodynamically consistent. We perform this matching across
a transition window from $n=6.3\times10^{-5}$ to 0.08~fm$^{-3}$. 
This ensures that at densities above 0.5$n_{\rm sat}$, we are exclusively using 
the approximate EoS model, but that at lower densities,
the full table is smoothly approached.

In order to compare the approximate and full versions of the 3D EoS tables in a 
merger simulation, there is one final step required:
the construction of the initial data. We describe this further in Sec.~\ref{subsec:ID}; but here,
we note that this additionally requires the extraction of a 1D cold, \beq~slice
from the approximate 3D EoS table. This, in turn, requires providing
chemical potentials in a way that is consistent with the rest of 
the $M^*$-approximation, in particular with the symmetry energy description.
We calculate approximate chemical potentials for the approximate SFHo table following
the procedure described in Appendix B of \cite{Most2021}.  A comparison of the resulting, approximate EoS
with the full version of SFHo is presented in Appendix~\ref{sec:EoSappendix}.

\section{Numerical methods}
\label{sec:methods}
In this section we highlight the numerical methods used to test our EoS 
framework. We include details on our extensions to the 
open-source \texttt{IllinoisGRMHD} code to allow for tabulated, 
finite-temperature EoS compatiblity. We also provide details on the 
initial conditions used for our simulations. Finally, we discuss the 
diagnostics used to compare and contrast the results of our simulations 
using the EoS tables described in Sec.~\ref{sec:EoS}.

\subsection{Evolution code}
\label{subsec:evol_code}
Our evolution code consists of an updated version of 
the \texttt{IllinoisGRMHD} code \citep{Etienne2015},
which is described and validated in detail in \citet{Espino2022}.
\texttt{IllinoisGRMHD} solves the equations of general relativistic 
ideal magneto-hydrodynamics (GRMHD) in a dynamical spacetime, within the 
Baumgarte-Shapiro-Shibata-Nakamura (BSSN) 
formulation~\citep{BSSN1,Shibata1995,Baumgarte1999} of the 3+1 
Arnowitt-Deser-Misner (ADM) formalism.
\texttt{IllinoisGRMHD} works with the spacetime metric
\begin{equation}
ds^2 = -\alpha^2 dt^2 + \gamma_{ij}(dx^i + \beta^i dt)(dx^j + \beta^j dt),
\end{equation}
where $\alpha$ is the lapse, $\beta^i$ is the shift, 
$\gamma_{\mu\nu} = g_{\mu\nu} + n_\mu n_\nu$ is the 
induced metric, and $n^\mu=(1/\alpha, \beta^i/\alpha)$ is the 
future-pointing unit vector orthogonal to each 
space-like hypersurface. 
Our updates to \texttt{IllinoisGRMHD} include the evolution
of the electron fraction $Y_{\rm e}$ and the use of state-of-the-art 
conservative-to-primitive routines which are compatible with finite-temperature
EoSs. Specifically, we supplement 
\texttt{IllinoisGRMHD} with the addition of the equation for $Y_{\rm e}$ 
advection, assuming conservation of charged lepton number,
\begin{equation}\label{eq:Ye_advec}
\partial_t(\tilde{Y}_{\rm e}) + \partial_j(v^j \tilde{Y}_{\rm e}) = 0,
\end{equation}
where 
$\tilde{Y}_{\rm e} \equiv \alpha Y_{\rm e} \sqrt{\gamma} \rho_{\rm b} u^0$, 
$\rho_{\rm b}$ is the rest mass density,
$\gamma$ is the determinant of the 3-metric, 
and $u^0$ is the temporal component of the fluid 4-velocity.
We emphasize that, as we assume the conservation of \emph{charged} lepton number,
Eq.~\eqref{eq:Ye_advec} only captures the advection of $Y_{\rm e}$. 
In the presence of neutrinos, we expect source terms to appear on the 
right-hand-side of Eq.~\eqref{eq:Ye_advec}, which can alter the
evolution of $Y_{\rm e}$~\citep{Radice:2016dwd, Foucart:2016rxm, Most2019, 2020zndo...3689751G, Radice:2021jtw}.
We leave the investigation of neutrino transport effects, with
the use of our approximate EOS tables, to future work.
The conservative-to-primitive routines within 
the public version of \texttt{IllinoisGRMHD}  
assume a polytropic, barotropic form for the EoS. To 
allow for generic, finite-temperature EoSs, we have added the 
conservative-to-primitive inversion algorithm of~\cite{Palenzuela_2015} 
to \texttt{IllinoisGRMHD}. This algorithm was originally 
implemented in the open-source conservative-to-primitive 
driver code of~\cite{Siegel:2017sav}, which we have adapted to the Cactus 
framework within which \texttt{IllinoisGRMHD} operates. The 
conservative-to-primitive algorithm of~\cite{Palenzuela_2015} as 
implemented in~\cite{Siegel:2017sav} provides a robust and efficient 
method for general conservative-to-primitive inversion when using 
tabulated, finite-temperature EoSs and has been used in several GRMHD 
codes~\citep{Most2019, 2020zndo...3689751G}.

The convergence of the updated \texttt{IllinoisGRMHD} code with a
similar tabulated EoS was recently studied in \cite{Espino2022}, where 
it was shown that the code converges at the expected second order rate,
and that the code's convergence properties are generally consistent with
those of other open-source GRMHD codes. We also find merger dynamics 
which are consistent with other open-source GRMHD codes. Importantly, 
our code produces similar merger times and remnant thermal profiles to 
the {\tt GRHydro}~\citep{2014CQGra..31a5005M}, 
{\tt Spritz}~\citep{2020zndo...3689751G}, 
and {\tt WhiskyTHC}~\citep{Radice_2014} codes
for astrophysical systems relevant to the present work (i.e., BNS mergers with
the use of finite-temperature EoS tables). We discuss the convergence 
properties of our code, and provide comparisons to the results produced 
by other codes in~\cite{Espino2022}; we refer the reader to that work 
for further details.

We evolve the spacetime using the {\tt McLachlan} spacetime evolution 
code~\citep{Brown:2008sb,Reisswig_2011GWs} within 
the {\tt EinsteinToolkit}~\citep{Loffler2012ETK}, which solves 
the Einstein 
equations within the BSSN formulation of the ADM 3+1 formalism.  
We evolve using the ``1+log" slicing condition for the 
lapse~\citep{Bona:1994dr} and a ``Gamma-driver" condition for the shift, 
with the shift coefficient set to $\nu=0.75$~\citep{Alcubierre:2002kk}. 
For time-integration, we use a fourth-order Runge-Kutta scheme with a 
Courant-Friedrichs-Lewy (CFL) factor of 0.5, provided by the {\tt MoL} 
thorn within the {\tt EinsteinToolkit}.

\subsection{Initial conditions}
\label{subsec:ID}
We construct binary neutron star initial data using the \texttt{LORENE} 
libraries.\footnote{https://lorene.obspm.fr/}
\texttt{LORENE} requires the use of cold, barotropic 
EoS tables corresponding to nuclear matter in neutrinoless \beq, 
such that the pressure 
$P_\beta=P_\beta(\rho_{\rm b})$. 
We extract such EoS slices from the 3D tables described in 
Sec.~\ref{sec:EoS} as follows: at each value of $\rho_{\rm b}$ available 
in the table, we fix the temperature to $T_{\rm cold}=0.1$~MeV and locate 
the table entry that corresponds to \beq, such that
\begin{equation}\label{eq:beta_equil}
\mu_{\rm n} - \mu_{\rm p} - \mu_{\rm e} = 0,
\end{equation}
where $\mu_{\rm n,p,e}$ are the neutron, proton, and electron
chemical potentials, respectively. At each value of $\rho_{\rm b}$ considered, 
Eq.~\eqref{eq:beta_equil} is satisfied for a unique value of 
$Y_{\rm e}=Y_{\rm e}^\beta$. This procedure allows us to 
extract barotropic tables 
corresponding to \beq, such that  
$P_\beta (\rho_{\rm b}) = P(\rho_{\rm b}, Y_{\rm e}^\beta, T_{\rm cold})$, 
where $P$ corresponds to the pressure in the full 3D table.

We use the resulting cold, \beq~slice of 
the approximate SFHo table in order to 
construct initial data for
an irrotational, equal-mass 
binary neutron star system on a quasi-circular orbit.
We use this initial data for both
the simulation with the full SFHo EoS, and the simulation with the approximate SFHo EoS. That is, we launch both simulations
from initial data that were constructed with the approximate version
of the SFHo table. Because the approximate version of the EoS
is designed to be identical to the full version of SFHo at zero-temperature
(as shown schematically in Fig.~\ref{fig:construction}),
the initial conditions should, in principle, be identical for both evolutions.
In practice, however, there are some small differences in
the approximate chemical potentials, which lead to small differences
in the extracted \beq~slices. As a result, the cold, \beq~EoSs
differ slightly; but, as we show in
Appendix~\ref{sec:EoSappendix}, the difference 
between the cold, \beq~pressures for the approximate and full versions of 
SFHo is $\lesssim$1\% at densities above half the nuclear saturation 
density. As a result, our choice to adopt identical initial 
data for both simulations has 
negligible impact on the evolutions, as we 
will demonstrate below.

The centers-of-mass of each binary component 
are initially separated by 45~km and each star has a baryonic mass 
of $M_{\rm b}=1.42$~$M_\odot$ and a gravitational mass of $M=1.26$~$M_\odot$,
where the gravitational mass is defined as the ADM mass of a static 
Tolman-Oppenheimer-Volkoff (TOV)~\citep{Tolman1939, Oppenheimer1939} star with the same 
baryonic mass. The total ADM mass of the system is $M_{\rm ADM}=2.59$~$M_\odot$.  
The central value of the specific 
enthalpy for each star is $h\approx 0.2112$, and total system angular 
velocity and orbital frequency are $\Omega\approx 1741$~rad/s and 
$f\approx 277$~Hz, respectively. 

For each binary evolution, we use seven spatial refinement levels, which
are separated by a 2:1 refinement ratio. The resolution on the finest
refinement level corresponds to a grid spacing of $\Delta x =152$~m,
which means there are $\sim128$~points across the diameter
of each initial neutron star.

\subsection{Diagnostics}
\label{sec:diagnostics}

We use several diagnostics to assess the state of our simulations and to
compare results between the full and approximate versions of the SFHo 
EoS. All codes used for diagnostics are available within the 
{\tt EinsteinToolkit} suite of codes~\citep{Loffler2012ETK}; in the 
following we highlight 
specific thorns used for each diagnostic, where relevant.

We monitor global scalar variables, including the maximum of the rest mass 
density, $\rho_{\text{b, max}}$, and the minimum of the lapse, 
$\alpha_{\rm min}$, to monitor partial gravitational collapse during the
merger. To visually assess thermal features of the merger, we consider 
2D 
equatorial snapshots of fluid variables, including the ratio of the 
thermal to 
the cold fluid pressure, $P_{\rm th}/P_{\rm cold}$, 
as well as the temperature, $T$. We also 
consider density profiles of these variables, to understand in a deeper 
manner the differences in thermal properties between 
the two simulations.

We extract gravitational waves within the Newman-Penrose (NP) 
formalism~\citep{Newman:1961qr, Penrose:1962ij}, 
by calculating the NP scalar $\Psi_4$ which is decomposed into $s=-2$ 
spin-weighted spherical harmonics, with use of the {\tt Multipole} thorn
within the {\tt EinsteinToolkit}. The coefficients of the decomposition 
are labeled as $\psi_4^{\ell,m}$. For the dominant $\ell=m=2$ mode,
we compute $|\psi_4^{2,2}(t)|$ at several extraction radii and 
report the value in the wave zone, and we use this to
extract the polarizations of the GW strain $h$,
\begin{equation}
\Psi_4 = \ddot{h}_+ - i \ddot{h}_\times,
\end{equation}
using the fixed-frequency integration method \citep{Reisswig2011}.

\label{subsec:diagnostics}

\section{Results}
\label{sec:results}

We now turn to the results of merger evolutions using the full and 
approximate 
SFHo tables.  
We start with an overview of the merger dynamics and
remnant properties, and then discuss in detail the thermal
properties of the remnant, in order
to understand how the EoS modeling affects the post-merger evolution.
Finally, we compare the gravitational wave emission for both
evolutions, as this is a directly observable signal which is sensitive
to the high-density EoS.
For all diagnostics, we find strong agreement between the results found using
the full and the approximate versions of the SFHo table.

 \begin{figure}
\centering
\includegraphics[width=0.4\textwidth,clip]{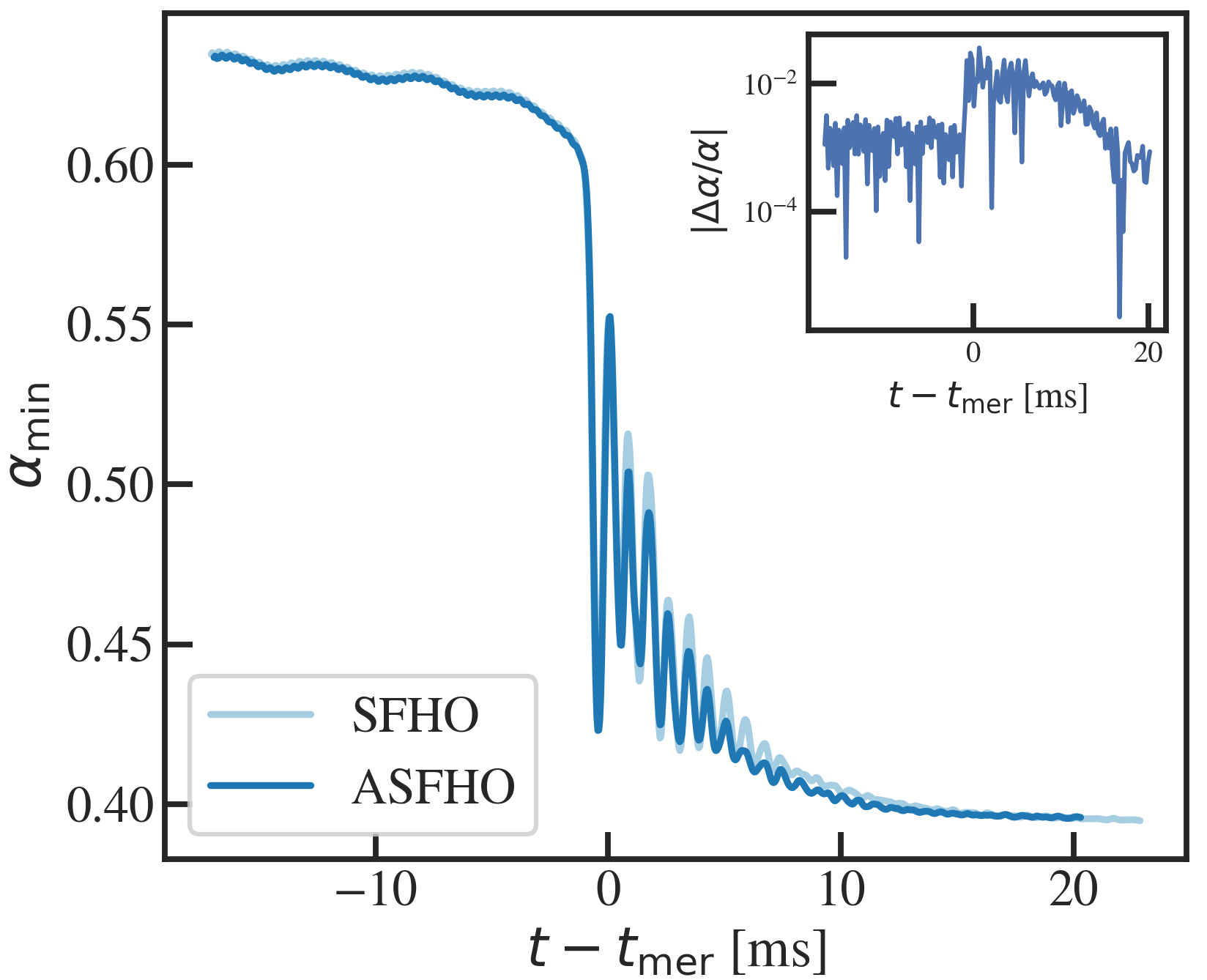}  \\
\includegraphics[width=0.4\textwidth,clip]{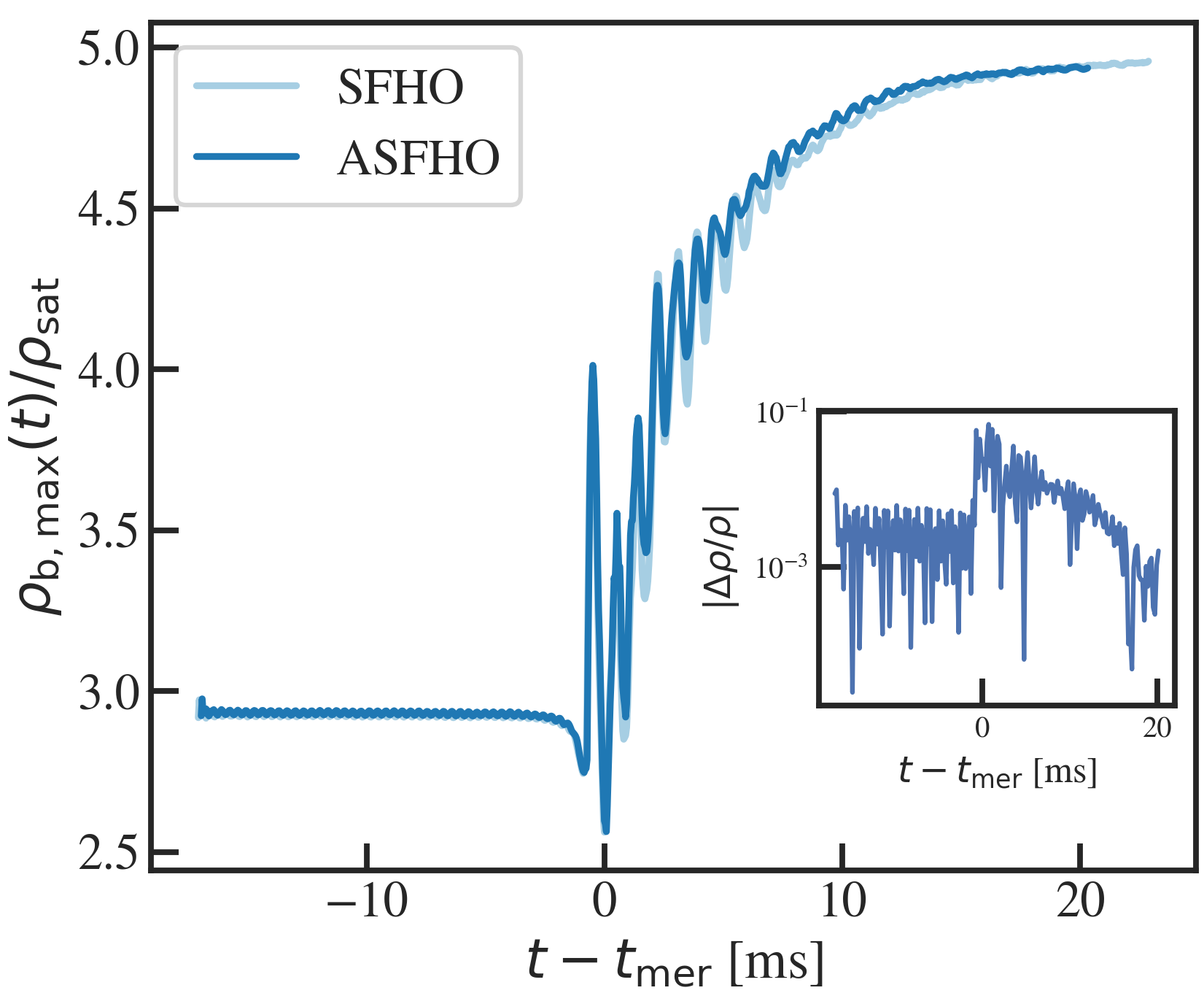} 
\caption{ \label{fig:rho_1D} Top: Minimum lapse as a function of time
for both evolutions. Bottom: Maximum rest mass density for the same evolutions,
normalized to the nuclear saturation density.
The results from the simulation using the full SFHo table are shown in light blue,
while the results with the approximate SFHo table (labeled ``ASFHO")
are shown in dark blue. The insets show the fractional difference between
the two cases.}
\end{figure}

 \begin{figure*}
\centering
\includegraphics[width=0.9\textwidth]{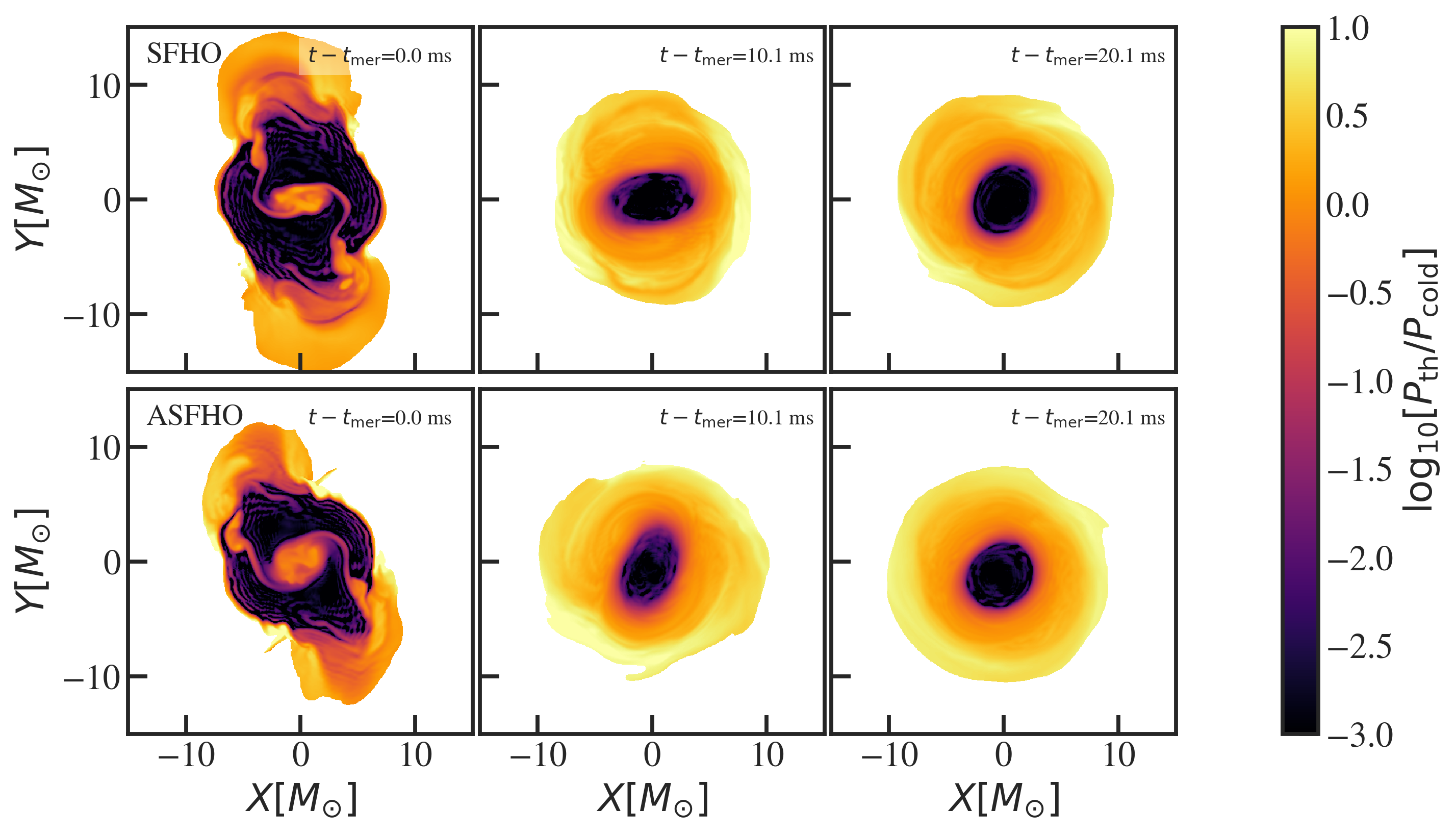}
\caption{ \label{fig:PthoPc_2D} 2D equatorial snapshots of the thermal pressure relative to the 
cold pressure, at three times  during and after the merger. The top row shows the results from
the simulation run with the full SFHo table, while the bottom row corresponds to the results from the evolution 
with the approximate SFHo table, constructed using the $M^*$-framework. These figures
include only matter with densities above $0.01\rho_{\rm b,max}(t=0)$; matter at lower
densities is masked in white. There is a slight difference in 
phase at merger due to the difference in initial conditions, but
we find strong agreement in the degree of heating and the spatial distribution of $P_{\rm th}/P_{\rm cold}$ 
with the approximate model compared to the full table for SFHo.}
\end{figure*}

\subsection{Merger overview}

We start with a short overview of the merger. For both evolutions, 
we track the final $\sim$6 orbits prior to merger.
Throughout the inspiral, we find that the neutron stars remain stable 
and show no signs of significant heating. In both cases, the
 binaries have a time-to-merger of 17.2~ms, (where this
time, $t_{\rm mer}$, is defined as the time when the gravitational wave
strain reaches a maximum; see Sec.~\ref{sec:GW}). The time-to-merger
is nearly identical for the two versions of the EoS, with only a 0.04~ms difference
 (0.2\% fractional difference) in $t_{\rm mer}$.

The rest mass of the merger remnant is $\sim2.88~\Ms$ in both cases, which
exceeds the maximum rest-mass of the zero-temperature Kepler
sequence of 2.85~$\Ms$ for this EoS. This suggests
that the remnant is likely supported by a combination of differential 
rotation and thermal pressure \citep{Paschalidis2012}, but that it will
eventually collapse. The timescales on which 
differential rotation is removed are significantly longer than our full 
evolution timescale~\citep{Paschalidis2017, 2021MNRAS.503..850I}; as such, 
we do not expect to find a black hole remnant during our 
simulations. Indeed, we find no signs of collapse by the end of our
simulations, which last $\sim$20~ms
past the merger. The top panel of Fig.~\ref{fig:rho_1D} 
shows the minimum lapse function
over the course of the evolution. It is approximately constant at late times,
indicating that the remnant is indeed stable for both evolutions.  

We additionally find strong agreement in the evolution of the maximum rest-mass
density between the two EoSs, as shown in the bottom panel of Fig.~\ref{fig:rho_1D}.
For both versions of the EoS, the maximum rest mass density
at the end of the evolutions is 4.94~$\rns$, with a fractional difference between
the two cases of only 0.16\%.

\subsection{Thermal properties of the remnant}
\label{sec:thermal}

 \begin{figure*}
\centering
\includegraphics[width=0.9\textwidth]{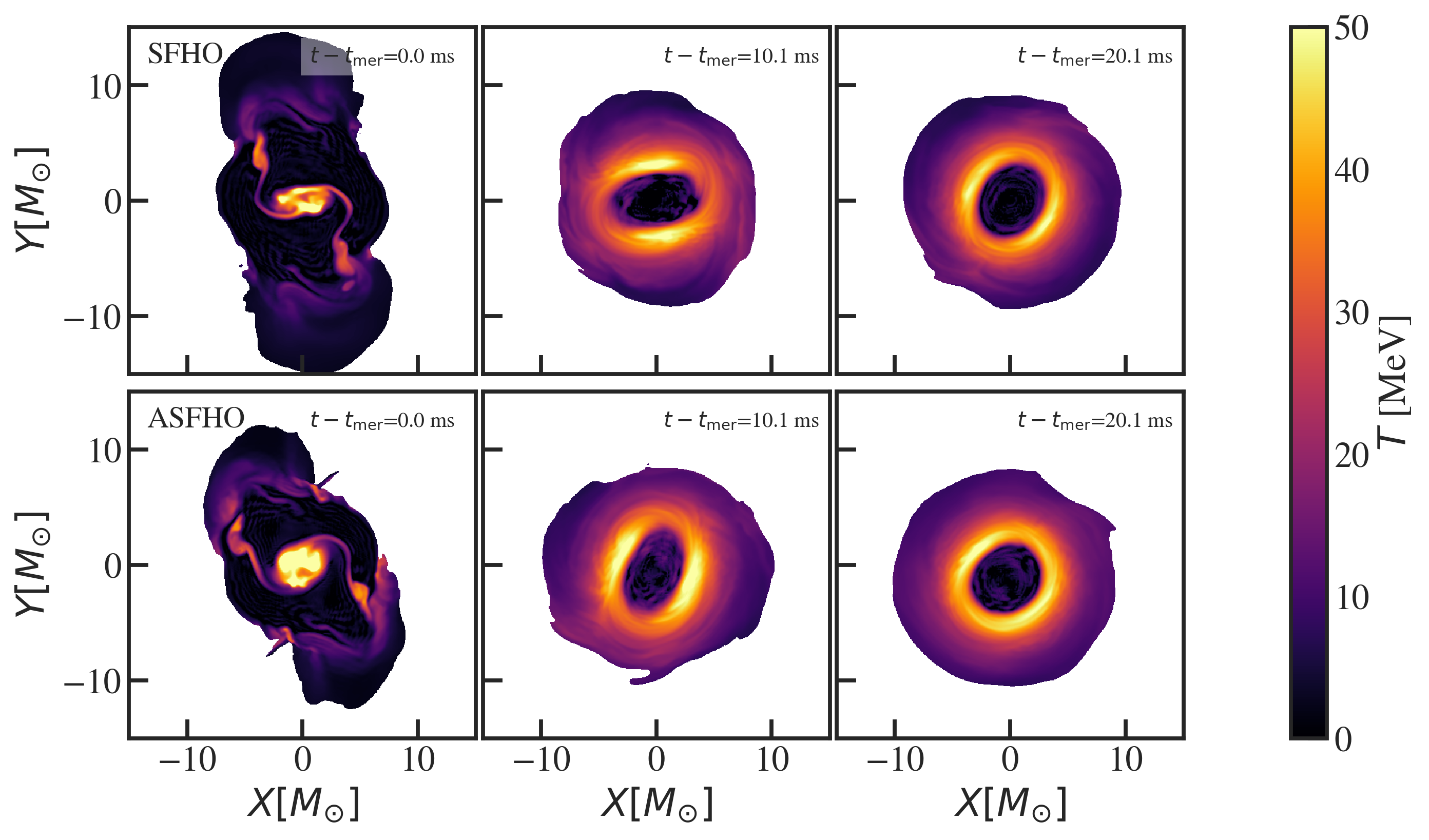}
\caption{ \label{fig:temp_2D} 2D equatorial snapshots of the temperature, at three times during and after the merger.
 The formatting of this figure is otherwise identical to Fig.~\ref{fig:PthoPc_2D}.}
\end{figure*}

The thermal profiles of the remnant are also very similar. We show equatorial snapshots of the thermal pressure,
relative to the cold pressure, in Fig.~\ref{fig:PthoPc_2D}. The results using the full SFHo table are
shown in the top row, and those from the evolution with the approximate SFHo table are
shown in the bottom row. The columns show snapshots at three times 
during and after merger.

From these snapshots, we find that
the degree of heating during and after merger is qualitatively very similar for both the full and 
approximate SFHo evolutions. In both cases, we find that there is strong shock
 heating at the merger interface,
which leads to substantial thermal pressure ($\gtrsim10\%$ of the cold pressure) 
in the outer layers of the stars. However, in the innermost 
core of the remnant, the thermal pressure remains subdominant to the 
cold pressure for both cases.

The corresponding temperatures for each snapshot are shown in Fig.~\ref{fig:temp_2D}.
We find peak temperatures in excess of 50~MeV for both versions of the EoS.
 Additionally, we find that the temperature peaks away from the center of the
  remnant (as seen in the bright ``ring" 
in the right columns of Fig.~\ref{fig:temp_2D}), which is a result of the particular
 $M^*$-parameters used here. In \citet{Raithel2021}, it was shown that the 
 depth to which the shock-heating penetrates the remnant depends on the 
 choice of $M^*$-parameters, with some choices of $M^*$-parameters leading to
 the appearance of a high-temperature ``ring" and other choices leading to temperatures
 that peak closer to the center of the remnant. The similarity of the
 temperature ring features in Fig.~\ref{fig:temp_2D} thus already starts to indicate that
 the thermal prescription in the $M^*$-version of the EoS provides a realiable
 approximation to that of the full SFHo table.

\begin{figure}
\centering
\includegraphics[width=0.45\textwidth]{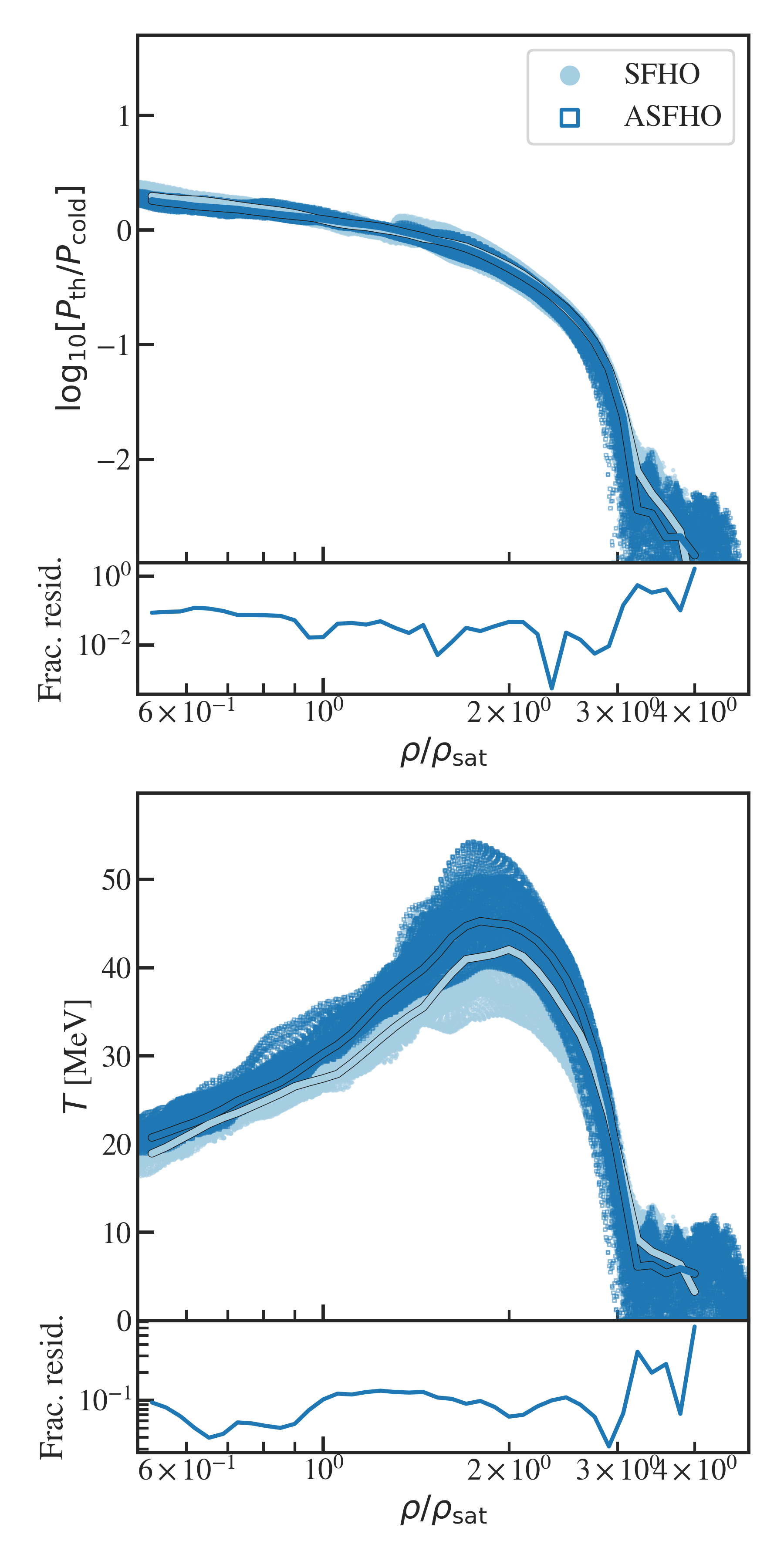}
\caption{ \label{fig:PthoPc_1D}  Late-time profiles of the thermal pressure and temperature
as a function of the density. The points illustrate the spread of values found on the simulation grid, 
while the lines indicate the median value, calculated within density bins that are log-uniformly sampled. 
The lighter lines/points correspond to the results from the simulation with full SFHo table, 
while the darker lines/points correspond to
the results from the approximate SFHo table (``ASFHO").
The lower panel to each figure shows the fractional residuals in the median values
of $P_{\rm th}/P_{\rm cold}$ and the median temperature, respectively.} 
\end{figure}

We explore these thermal profiles more quantitively in Fig.~\ref{fig:PthoPc_1D}, where we show
each quantity as a function of the density for the final snapshot ($t-t_{\rm mer}\sim$20~ms).
In this figure, the markers represent values of a particular grid
point in our simulations, while the lines
represent the median of all values, calculated within
density bins that are log-uniformly spaced. The density profiles of $P_{\rm th}/P_{\rm cold}$
are shown in the top panel of Fig.~\ref{fig:PthoPc_1D}, while the temperature profiles
are shown in the bottom panel.

We find that the thermal pressure profiles differ only slightly between the full 
and approximate versions of SFHo. In particular, for $\rho \approx 1-3 \rns$, the
difference in the median value of $P_{\rm th}/P_{\rm cold}$  is $\lesssim5\%$ 
between the two versions of the EoS. In both cases, we find that the thermal
pressure is a significant fraction of the cold pressure,
with $P_{\rm th} \gtrsim 0.1 P_{\rm cold}$ for densities below 3 $\rns$. 
Overall, this suggests a high degree of heating that extends to high
densities within the star, 
and that the $M^*$-framework is able to accurately capture this response.

We likewise find close agreement in the temperatures reached in each simulation,
with the approximate version of SFHo leading to slightly higher 
temperatures in the late-time remnant, but with a very similar density-dependence.
 In particular, Fig. \ref{fig:PthoPc_1D} shows that the peak 
 of the median temperature profile occurs at densities of $\sim2\rns$ for both EoSs. 
For the evolution with the full version of SFHo, the median temperature profile
 in Fig. \ref{fig:PthoPc_1D} peaks at a value of 42~MeV, whereas 
 the median temperature profile for the evolution with the approximate EoS peaks at 45~MeV
 (fractional difference of 6\%).
In general, for $\rho<3\rns$, we find that the median temperatures agree to within 
$\lesssim$10\% between the two EoSs. 

To place these differences in context, we can also consider the extent to which 
$P_{\rm th}/P_{\rm cold}$ and the temperature vary purely by changing the choice
of $M^*$-parameters. In \cite{Raithel2021}, an extremal range of $M^*$-parameters
was explored in a series of $1.4+1.4\Ms$ neutron star mergers, 
all governed by the same cold EoS (ENG).
The choice of $M^*$-parameters was designed to bracket the range of uncertainty spanned
by a sample of nine existing finite-temperature EoS tables. In that work, it 
was found that the median $P_{\rm th}/P_{\rm cold}$ can vary by a factor
of a few shortly after merger, while the median temperature can vary
by $\sim2\times$, for the extreme set of $M^*$-parameters.  Thus, when 
compared to the range of outcomes allowed by the degrees of freedom of the full $M^*$-model,
the agreement between the full and approximate SFHo tables is very strong. 
 We note one caveat when comparing to the results of \cite{Raithel2021}: namely, that
 the exact range of thermal outcomes depends not only on the $M^*$-parameters explored,
 but also on the cold EoS, as well as the binary mass and mass ratio of the merger. 
 Additional parameter studies to quantify these dependencies will be the subject of future work.

To summarize the results of this section: we find only small differences in the
thermal pressure and temperature profiles when using the approximate version of the
EoS. Moreover, we find that the qualitative shape and density-dependence of
these profiles is accurately recreated with the $M^*$-framework.

\subsection{Gravitational wave signal}
\label{sec:GW}
\begin{figure}
\centering
\includegraphics[width=0.45\textwidth, clip]{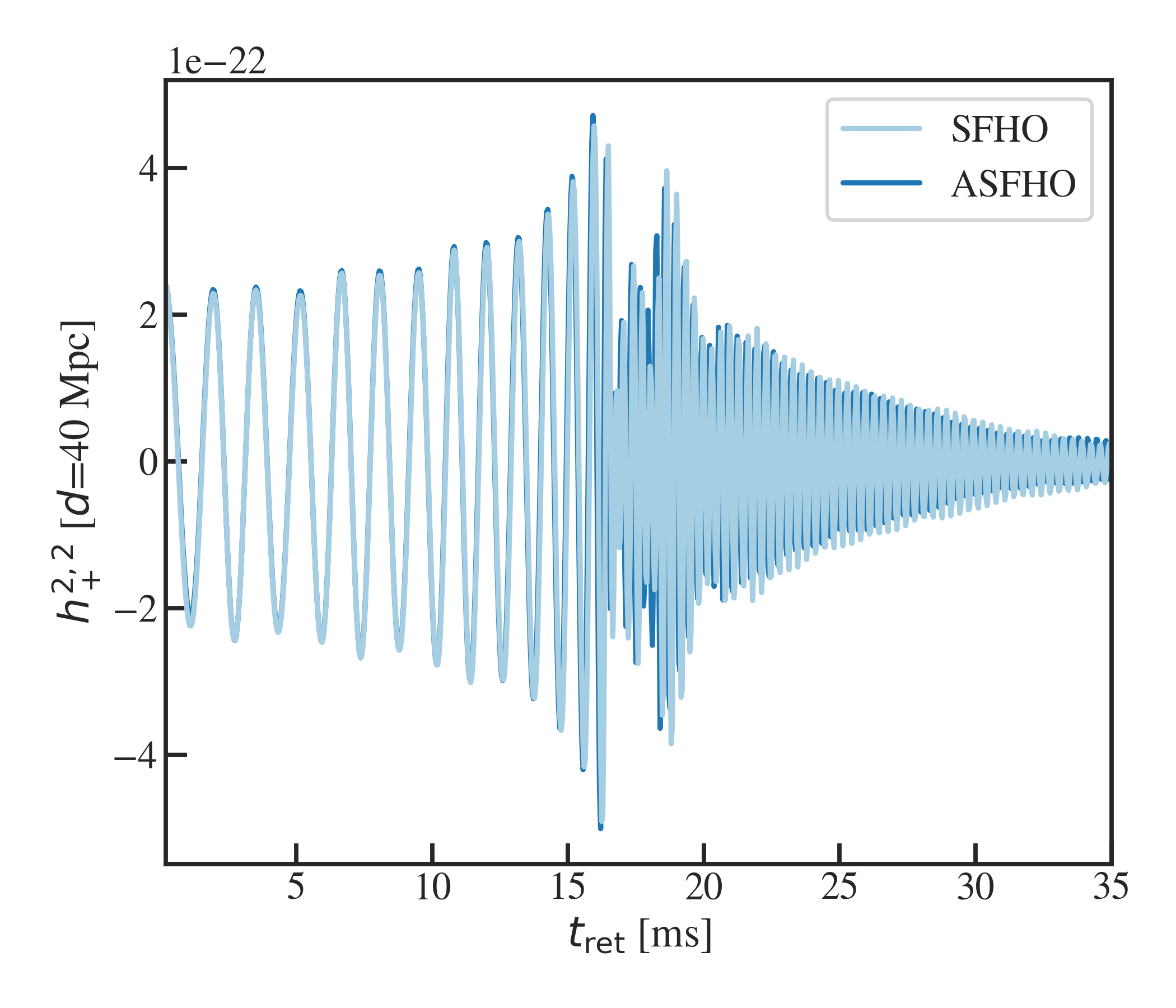} 
\caption{ \label{fig:strain} The $\ell=m=2$ mode of the plus-polarized gravitational wave strain
for a face-on merger located at a distance of 40~Mpc. The strain is plotted as a function
of the retarded time, with the evolution using the full SFHo table shown in light blue,  and the evolution
using the approximate table shown in dark blue. }
\end{figure}

 For the remainder of the paper, we turn now to the gravitational wave (GW) signal,
 as the key observable signature which is sensitive to the high-density
 part of the EoS.
 In Fig.~\ref{fig:strain}, we show the $\ell=m=2$ modes of the
 plus-polarized gravitational wave strain for
 each simulation,
 which have been computed as described in Sec.~\ref{sec:diagnostics}.
 This signal corresponds to a face-on merger located at 40~Mpc. 
 We find that the inspiral portion of the GWs are nearly identical, as
 expected based on the fact that the zero-temperature, \beq~slices of the EoSs
 are also nearly identical (to within $\lesssim$1\% accuracy; see Appendix~\ref{sec:EoSappendix})
 and the fact that the neutron stars remain thermodynamically cold 
 ($P_{\rm th} \ll P_{\rm cold}$) during the inspiral.

In contrast, as discussed above, there is 
significant heating at and following merger, such that the pressure profile
of the post-merger remnant has a substantial ($\gtrsim10\%$)
contribution from $P_{\rm th}$ at densities up to $3\rns$.
Thus, if any differences were to arise between the GWs
 with the full and approximate versions of SFHo, we expect they would be 
most evident in the post-merger phase.
Correctly modeling this phase of GW emission is of particular 
importance, as many studies have found empirical correlations
between the oscillation frequencies of the post-merger remnant
and properties of the neutron star EoS, such as the radius,
stellar compactness, or tidal deformability (e.g.,
\citealt{Bauswein2012,Bauswein2012a,Takami2014,Bernuzzi2015,Vretinaris2020};
see also \citealt{Raithel2022}). In principle, such correlations make it possible
to use a measurement of post-merger GWs to place immediate
constraints on the neutron star EoS. 

 \begin{figure}
\centering
\includegraphics[width=0.45\textwidth, clip]{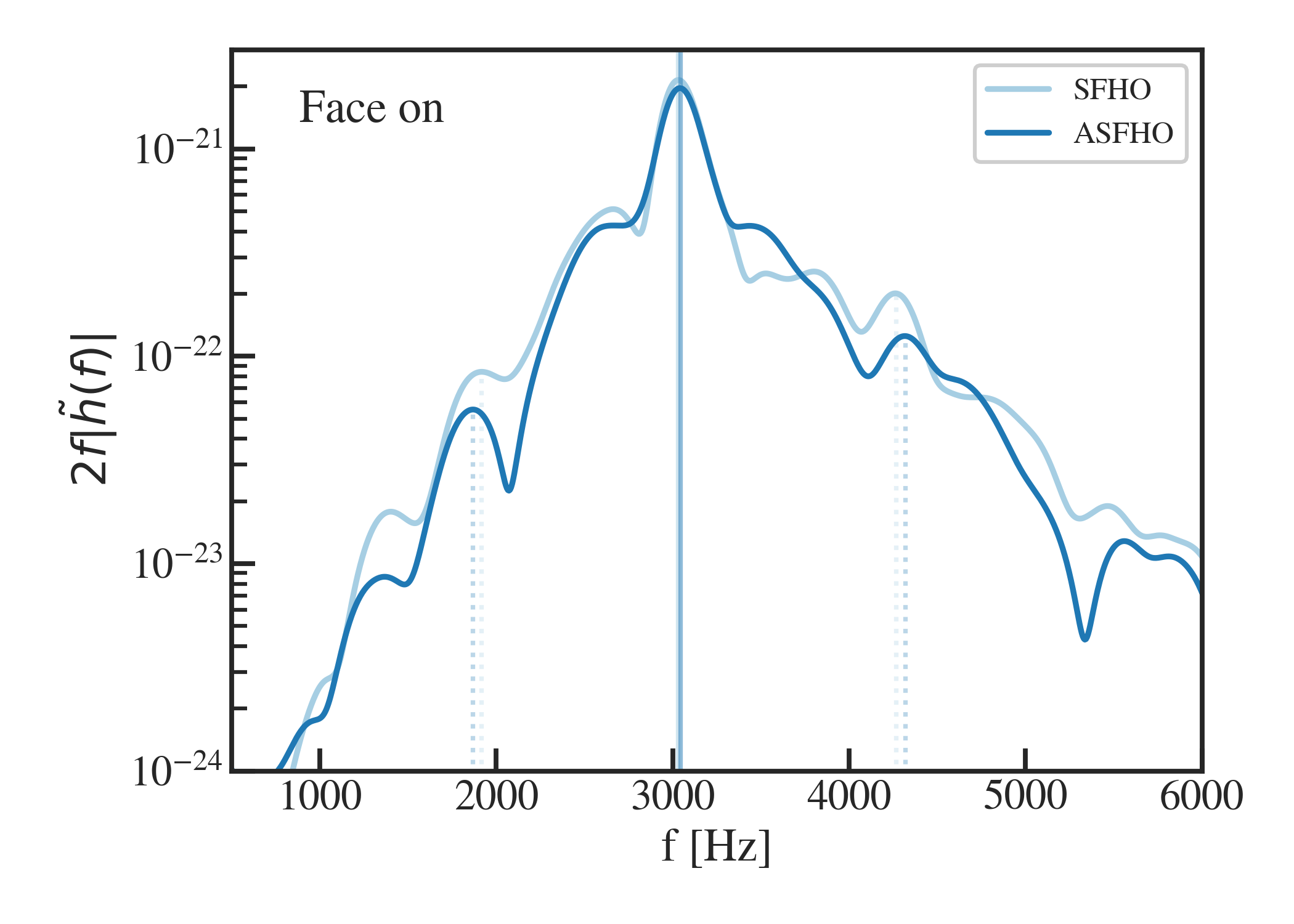} \\
\includegraphics[width=0.45\textwidth, clip]{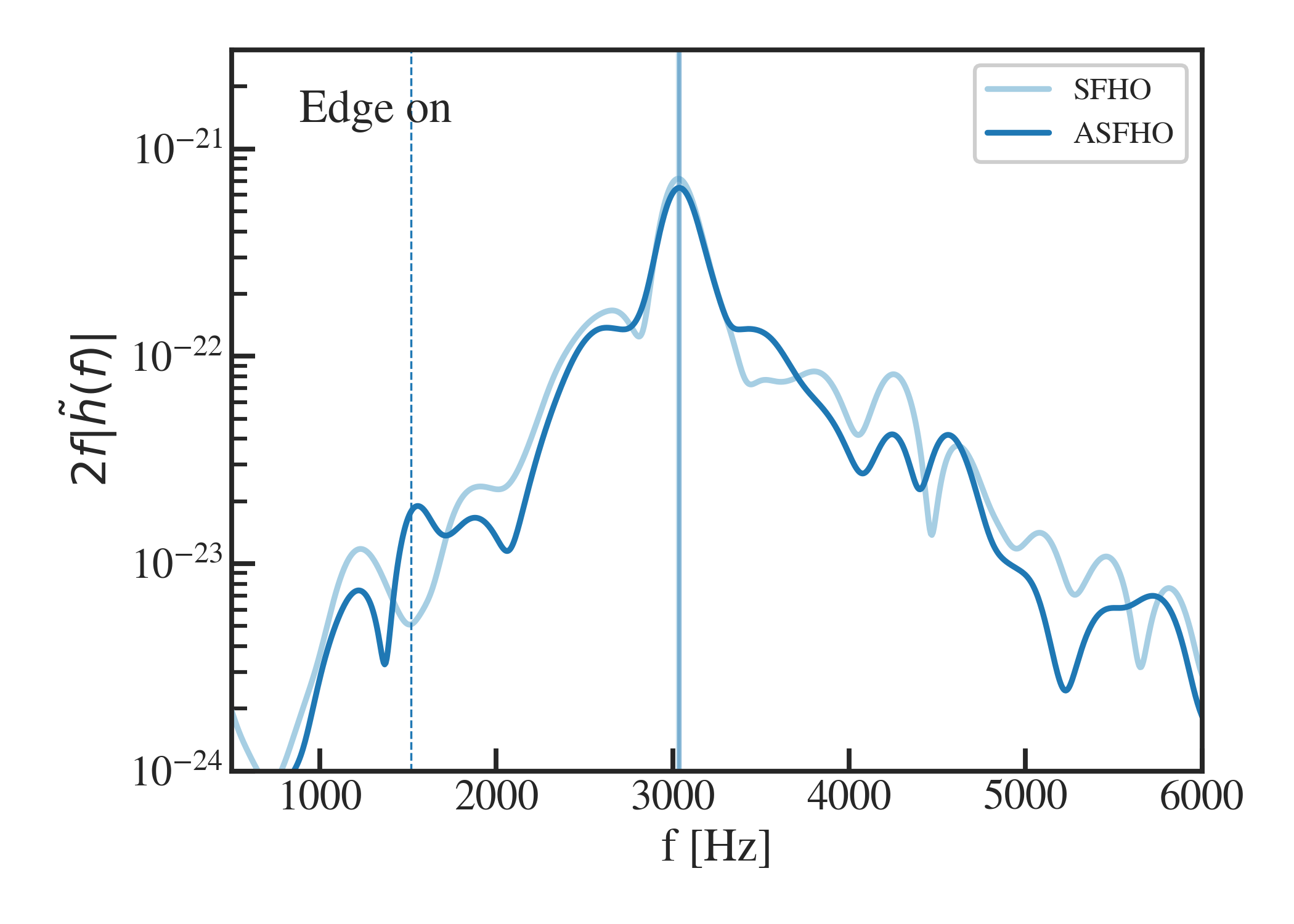}
\caption{ \label{fig:hc}  Characteristic strain of the post-merger
gravitational wave signal. These spectra include all $\ell=2,3$ modes of the gravitational wave strain. 
The spectra for an optimally-oriented merger located at a distance of 40~Mpc are shown in the top panel,
while the bottom panel shows the spectra for an edge-on orientation.
The vertical lines indicate the location of the peak GW 
frequencies; the dotted lines (shown only for the face-on case) indicate the location
of the secondary GW peak frequencies; and the dashed lines (shown only for the edge-on case)
indicate the location of the $m=1$ spectral peak.}
\end{figure}

In order to explore these post-merger oscillation frequencies in more detail,
we calculate the characteristic strain, $2f|\tilde{h}(f)|$, where
$f$ is the frequency and $\tilde{h}(f)$ is the Fourier transform of the strain.
We compute this characteristic strain for the 20~ms signal immediately
following the merger, using six overlapping segments, which are
windowed and padded as described in Appendix C of \cite{Most2021}.
We show the resulting spectra, including all $\ell=2,3$ modes of the strain,
in Fig.~\ref{fig:hc}. The top panel in this figure shows the spectra
for an optimally-oriented (face-on) merger, while the bottom panel shows
the spectra for an edge-on configuration. The solid vertical lines in these
figures indicate the peak frequency of the gravitational waves,
 which we find to be 3.03~kHz for the evolution with the full version of the EoS,
 and 3.04~kHz for the evolution with the approximate version of SFHo. This difference
 of 12~Hz corresponds to a 0.4\% fractional difference between the two cases.
 We additionally identify two secondary peaks, which are located
 equidistantly above and below the dominant peak, at 
 $\sim$1.9 and 4.3~kHz. These are indicated with the dotted vertical 
 lines in the top panel of Fig.~\ref{fig:hc}, and agree between the full and approximate 
 versions of the EoS with fractional differences of $\sim$1-2\%.
 We note that the amplitude of these secondary peaks differ by $\sim$40\%; however,
 the majority of the power is in the main spectral peak, where both EoSs
 lead to similar amplitudes, to within 9\%.

 \begin{figure}
\centering
\includegraphics[width=0.45\textwidth, clip]{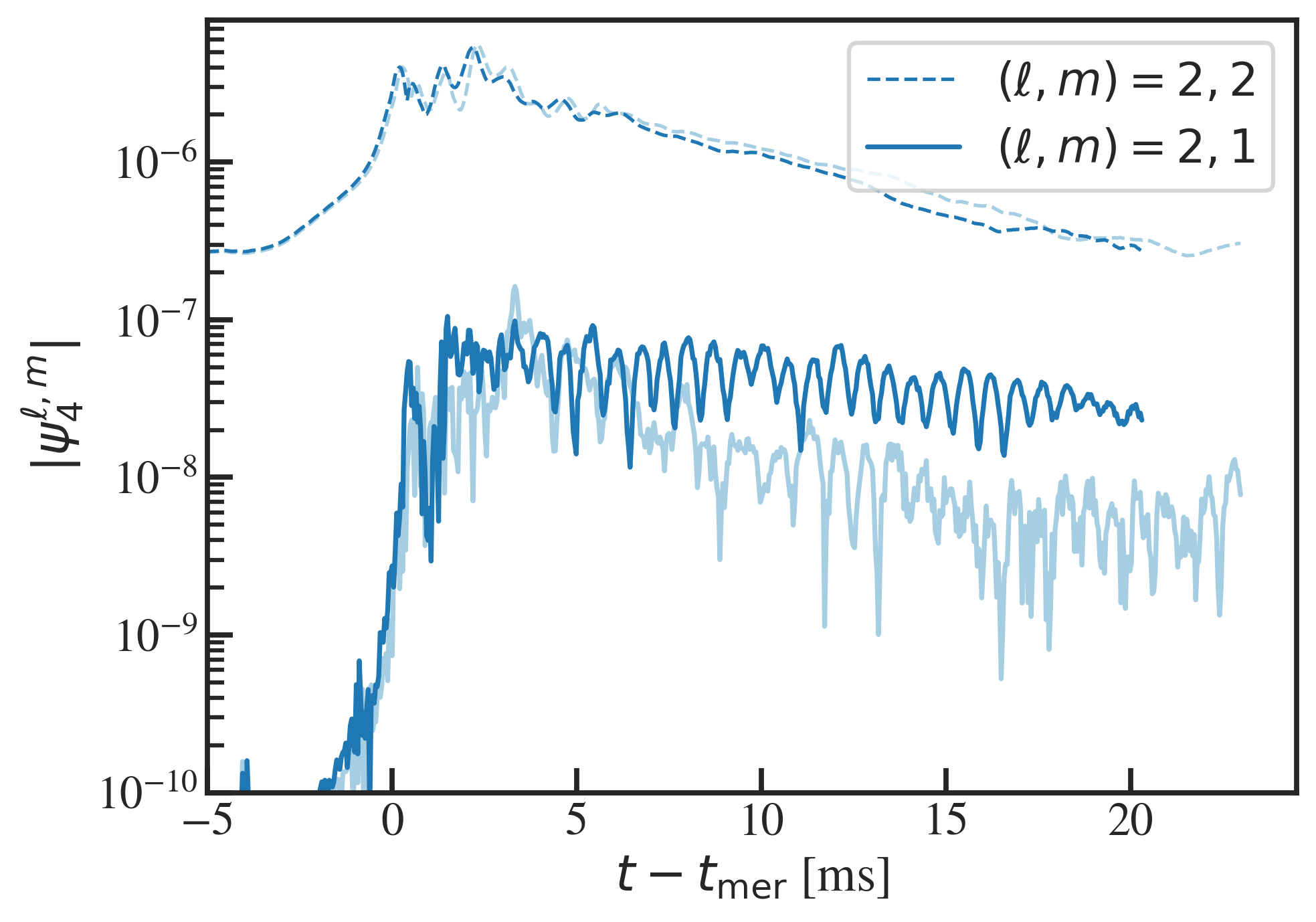} 
\caption{ \label{fig:Clm} Amplitude of the Newman-Penrose scalar coefficients, after decomposition
onto $s=-2$ spin-weighted spherical harmonics. The dashed lines show the $(\ell,m)=2,2$ mode
of $\psi_4$, while the solid lines represent the $(\ell,m)=2,1$ mode. The color-coding is the
same as the other figures in this paper, with the evolution using the full SFHo table
shown in light blue,  and the evolution using the approximate table shown in dark blue. }
\end{figure}

In the edge-on version of the spectra, we find a small but interesting difference
located at $f_{\rm peak}/2$, where $f_{\rm peak}\simeq3$~kHz is the peak frequency. A
feature at this location can indicate the formation and saturation of a one-arm,
spiral instability in the remnant, which drives
the production of $(\ell,m)=2,1$ gravitational waves
\citep{Paschalidis:2015mla,East:2015vix,East:2016zvv,Lehner:2016wjg,Radice:2016gym}.
In Fig.~\ref{fig:hc},
we find evidence of this $m=1$ spectral peak only for the evolution 
that uses the approximate version of the EoS table, but we find 
no clear $m=1$ peak in the evolution that uses the full SFHo table.

Interestingly, we find that the $m=1$ mode does indeed develop and 
saturate for both versions of the EoS,
but that it decays away slightly faster when the full version of the 
EoS table is used. This behavior is illustrated in Fig.~\ref{fig:Clm}, which shows
the coefficients of $\psi_4^{\ell,m}$ over time for the 
dominant $(\ell,m)=2,2$ mode, as well as for the $(\ell,m)=2,1$ 
mode in question. We find the evolution
of the dominant (2,2) mode is similar for both cases (consistent with the strains
shown in Fig.~\ref{fig:strain}), and that, in both cases,
there is a similar growth of the $m=1$ mode at merger. However, in the evolution
with the approximate version of the SFHo table, the $m=1$ mode stays saturated during
the evolution, whereas it starts to decay approximately
5~ms after the merger when the full version of the SFHo table is used.

This subtle difference in the decay rate of $\psi_4^{2,1}$ 
suggests that the $m=1$ mode may be somewhat more sensitive 
to the details of the thermal treatment, compared to the other spectral peaks discussed above.
Nevertheless, we emphasize that this difference between the two evolutions is small and that,
overall, we find strong agreement between the spectra near the dominant peak, 
which is expected to be the most easily observable feature for the near future.
 
 In summary, these results confirm that simulations performed with the $M^*$-framework
 are able to accurately reproduce the dominant post-merger spectral frequencies found with the full SFHo EoS,
 with $\lesssim1-2$\% errors introduced by the approximations of the $M^*$ model.

\section{Discussion and conclusions}
\label{sec:conclusion}
In this work, we have performed a  validation of the $M^*$-framework
for calculating parametric, finite-temperature EoSs in the context of binary neutron star mergers.
The framework, which was first introduced in \cite{Raithel2019}, 
was shown in that work to recreate the pressure at finite-temperatures 
and arbitrary electron fractions with errors of $\lesssim30\%$,
 for a sample of nine microphysical EoS tables.
In this work, we choose one of these EoS tables, SFHo, for which we
compare the outcomes of two full neutron star merger simulations, 
using either the full table or an approximate version, constructed
with best-fit $M^*$-framework parameters. The SFHo EoS provides
a challenging test case, as this is the softest of the original sample
of nine EoSs. As a soft EoS, it predicts small neutron stars which 
are expected to collide with high impact energies, leading
to significant shock heating. Additionally, due to the soft
(i.e., relatively shallow) pressure profile of the cold EoS, 
errors in the thermal pressure will be dynamically
more important than for the case of a stiffer cold EoS. 

We have shown that the approximate version of SFHo,
calculated with the $M^*$-framework, accurately captures the
merger dynamics and remnant properties that are found
with the full version of the EoS. In particular, we have compared the 
time-to-merger, rest mass of the remnant, and maximum rest-mass
density of the remnant, all of which agree with the values found
with the full EoS, with differences at the sub-percent level.

We have also investigated the thermal profiles of the remnant interior.
We find very good agreement in the density-dependence and magnitude of 
$P_{\rm th}/P_{\rm cold}$ for the two EoSs. We likewise find
good agreement in the temperature profiles, with the approximate
version of SFHo leading to marginally higher temperatures
in the late time remnant. Nevertheless, the median temperature in the late-time
remnant ($t-t_{\rm mer}\sim20$~ms) agrees for both versions of
the EoS to within $\lesssim$10\%, for $\rho<3\rns$. We additionally 
find that the $M^*$-framework accurately captures the density-dependence
of the temperature, with the temperature peaking at $2\rns$ in 
both cases, and decreasing at higher densities.

We place a particular emphasis on the gravitational wave signals
predicted by the merger simulations, as these are an important ingredient
for accurately interpreting upcoming observations.
We find strong agreement in the inspiral gravitational wave strains, as
expected based on the lack of significant heating prior to merger.
The stricter test comes in the post-merger phase, when the thermal pressure
exceeds 10\% of the cold pressure (for $\rho < 3\rns$) and thus can be
 dynamically important in influencing the post-merger evolution. It is
 in this regime that errors in the $M^*$-approximations will be
 most evident. Even for the ``stress test" of this soft EoS, 
 we find only small differences in the post-merger
 gravitational wave strains. In particular, the peak
 frequencies of the characteristic strain agree identically,
 to within the frequency resolution of our spectra. The
 secondary peak frequencies also agree with the results from the full EoS,
 to within $1-2\%$.
 
We note that the $M^*$-framework was 
  originally calibrated against a sample of EoSs that were calculated using relativistic
  energy density functionals, for which the model performed very well
  (e.g., errors of $\lesssim15\%$ in the $T=10$~MeV  thermal pressure
  above nuclear densities; \citealt{Raithel2019,ROPerratum}).
  One natural question, then, is  how general the framework and the 
  associated validation presented in this work may be.
  In \cite{Raithel2019}, the model was also compared against two 
  non-relativistic Skyrme energy functionals, as well as a two-loop model, 
  which is an extension of mean field theory. It was shown
  that the $M^*$-model had larger errors when compared against these 
  non-RMF EoSs, but that the $M^*$-model
  still offered a significant improvement over the hybrid approach 
  for a range of densities and temperatures.
  Recently, it was shown that the $M^*$-framework also
  performs very well when compared against calculations using nuclear many-body theory,
  with errors in pressure of $\lesssim6\%$ at supranuclear densities  \citep{Tonetto:2022zhs}. 
  In contrast, recent results from chiral effective field theory have shown that
  repulsive three-body interactions can 
  cause the effective mass function to start to rise at supranuclear densities
  \citep{Carbone:2019pkr,Keller:2020qhx}, which will modify the 
  density-dependence of the thermal pressure, and the effect of which is not included in
  the current $M^*$-model. We anticipate that the impact of 
  this modified density-dependence for $M^*$ would be smaller than, say,
  adopting a constant thermal index 
  (as has been explored e.g., in \citealt{Bauswein2010,Figura2020,Raithel2021}),
  but this would be interesting to explore in future work.

In summary, we find that the merger dynamics, remnant properties,
and gravitational waves
found with the $M^*$-framework closely
recreate the results found with the full version of the SFHo EoS.
This provides a strong numerical validation of the $M^*$-framework,
complementary to the analytic validation of the model reported
 in \ROP. More generally, these results confirm that the ($\lesssim$30\%)
 errors introduced into the analytic EoS by the approximations of the
 $M^*$-framework do not significantly affect binary neutron star merger outcomes.
 Rather, these errors lead to minor (typically percent or sub-percent)
 differences in the merger outcomes, for best-fit $M^*$ parameters.
 
 Of course, when exploring a new part of the EoS parameter space,
 the best-fit model parameters are not known a priori. These must be 
 bounded, either by experiment, theory, or by fits to existing samples of realistic EoSs. 
 Understanding how the uncertainties
 in these parameters affect the merger properties requires systematic parameter
 surveys. A first such study has already been performed in \cite{Raithel2021},
 where the authors bounded the range of merger outcomes for an extremal
 set of $M^*$-parameters, combined with a single, cold EoS.
 In that work, it was shown that by varying the $M^*$-parameters across
 an extremal range, the thermal pressure and temperature can vary significantly
 after merger. In comparison, the $\lesssim$5-10\% deviations in thermal 
 profiles that we find in this work between the
 the full and approximate SFHo tables are
 much smaller than the range of outcomes allowed by freely varying the $M^*$-parameters.
 While it may be possible to further
 improve this level of agreement 
 by finely-tuning the $M^*$-parameters, that is not the goal of this study.
 Rather, this study aims to address whether the general functional form of the thermal framework,
 parametrized at the level of the effective mass function, can recreate realistic merger
 outcomes for an optimal set of parameters. The fact that the approximate and full tables
 agree so well -- especially when compared to the range of outcomes allowed by the
 degrees of freedom of the $M^*$-model -- provides strong validation of the approach.
 Additional parameter
 surveys to expand on the results of \cite{Raithel2021} 
 will be the subject of future work.
 
 In the meantime, the present study confirms that the $M^*$-framework
 can be reliably used to probe neutron star merger properties in numerical
 simulations in full general relativity.

\section*{Acknowledgements}
CR would like to thank Elias Most for insightful
discussions related to this work.  CR gratefully acknowledges
support from a joint postdoctoral fellowship at the Princeton Center
for Theoretical Science, the Princeton Gravity Initiative, and as a
John N. Bahcall Fellow at the Institute for Advanced Study.  
PE acknowledges support from NSF Grant PHY-2020275
(Network for Neutrinos, Nuclear Astrophysics, and Symmetries (N3AS)).
This work
was in part supported by NSF Grant PHY-1912619 and PHY-2145421 to the
University of Arizona.  The simulations presented in this work were
carried out with the {\tt Stampede2} cluster at the Texas Advanced 
Computing Center and the {\tt Expanse} cluster at San Diego Supercomputer 
Center, under XSEDE allocation PHY190020. The simulations were also
performed, in part, with the Princeton Research Computing resources at
Princeton University, which is a consortium of groups led by the
Princeton Institute for Computational Science and Engineering
(PIC-SciE) and Office of Information Technology's Research
Computing. 

\section*{Data Availability}
The data underlying this article will be shared on reasonable request to the corresponding author.

\bibliography{gwthermal}

\begin{thebibliography}{}
\makeatletter
\relax
\def\mn@urlcharsother{\let\do\@makeother \do\$\do\&\do\#\do\^\do\_\do\%\do\~}
\def\mn@doi{\begingroup\mn@urlcharsother \@ifnextchar [ {\mn@doi@}
  {\mn@doi@[]}}
\def\mn@doi@[#1]#2{\def\@tempa{#1}\ifx\@tempa\@empty \href
  {http://dx.doi.org/#2} {doi:#2}\else \href {http://dx.doi.org/#2} {#1}\fi
  \endgroup}
\def\mn@eprint#1#2{\mn@eprint@#1:#2::\@nil}
\def\mn@eprint@arXiv#1{\href {http://arxiv.org/abs/#1} {{\tt arXiv:#1}}}
\def\mn@eprint@dblp#1{\href {http://dblp.uni-trier.de/rec/bibtex/#1.xml}
  {dblp:#1}}
\def\mn@eprint@#1:#2:#3:#4\@nil{\def\@tempa {#1}\def\@tempb {#2}\def\@tempc
  {#3}\ifx \@tempc \@empty \let \@tempc \@tempb \let \@tempb \@tempa \fi \ifx
  \@tempb \@empty \def\@tempb {arXiv}\fi \@ifundefined
  {mn@eprint@\@tempb}{\@tempb:\@tempc}{\expandafter \expandafter \csname
  mn@eprint@\@tempb\endcsname \expandafter{\@tempc}}}

\bibitem[\protect\citeauthoryear{Alcubierre, Bruegmann, Diener, Koppitz,
  Pollney, Seidel  \& Takahashi}{Alcubierre et~al.}{2003}]{Alcubierre:2002kk}
Alcubierre M.,  Bruegmann B.,  Diener P.,  Koppitz M.,  Pollney D.,  Seidel E.,
    Takahashi R.,  2003, \mn@doi [Phys. Rev.] {10.1103/PhysRevD.67.084023},
  D67, 084023

\bibitem[\protect\citeauthoryear{{Baiotti}}{{Baiotti}}{2019}]{Baiotti2019}
{Baiotti} L.,  2019, \mn@doi [Progress in Particle and Nuclear Physics]
  {10.1016/j.ppnp.2019.103714}, \href
  {https://ui.adsabs.harvard.edu/abs/2019PrPNP.10903714B} {109, 103714}

\bibitem[\protect\citeauthoryear{{Baiotti} \& {Rezzolla}}{{Baiotti} \&
  {Rezzolla}}{2017}]{Baiotti2017}
{Baiotti} L.,  {Rezzolla} L.,  2017, \mn@doi [Reports on Progress in Physics]
  {10.1088/1361-6633/aa67bb}, \href
  {http://adsabs.harvard.edu/abs/2017RPPh...80i6901B} {80, 096901}

\bibitem[\protect\citeauthoryear{{Baiotti}, {Giacomazzo}  \&
  {Rezzolla}}{{Baiotti} et~al.}{2008}]{Baiotti2008}
{Baiotti} L.,  {Giacomazzo} B.,   {Rezzolla} L.,  2008, \mn@doi [\prd]
  {10.1103/PhysRevD.78.084033}, \href
  {http://adsabs.harvard.edu/abs/2008PhRvD..78h4033B} {78, 084033}

\bibitem[\protect\citeauthoryear{{Baumgarte} \& {Shapiro}}{{Baumgarte} \&
  {Shapiro}}{1999}]{Baumgarte1999}
{Baumgarte} T.~W.,  {Shapiro} S.~L.,  1999, \mn@doi [\prd]
  {10.1103/PhysRevD.59.024007}, \href
  {https://ui.adsabs.harvard.edu/abs/1999PhRvD..59b4007B} {59, 024007}

\bibitem[\protect\citeauthoryear{{Bauswein} \& {Janka}}{{Bauswein} \&
  {Janka}}{2012}]{Bauswein2012a}
{Bauswein} A.,  {Janka} H.-T.,  2012, \mn@doi [Physical Review Letters]
  {10.1103/PhysRevLett.108.011101}, \href
  {http://adsabs.harvard.edu/abs/2012PhRvL.108a1101B} {108, 011101}

\bibitem[\protect\citeauthoryear{{Bauswein}, {Oechslin}  \& {Janka}}{{Bauswein}
  et~al.}{2010a}]{Bauswein2010a}
{Bauswein} A.,  {Oechslin} R.,   {Janka} H.~T.,  2010a, \mn@doi [\prd]
  {10.1103/PhysRevD.81.024012}, \href
  {https://ui.adsabs.harvard.edu/abs/2010PhRvD..81b4012B} {81, 024012}

\bibitem[\protect\citeauthoryear{{Bauswein}, {Janka}  \& {Oechslin}}{{Bauswein}
  et~al.}{2010b}]{Bauswein2010}
{Bauswein} A.,  {Janka} H.-T.,   {Oechslin} R.,  2010b, \mn@doi [\prd]
  {10.1103/PhysRevD.82.084043}, \href
  {http://adsabs.harvard.edu/abs/2010PhRvD..82h4043B} {82, 084043}

\bibitem[\protect\citeauthoryear{{Bauswein}, {Janka}, {Hebeler}  \&
  {Schwenk}}{{Bauswein} et~al.}{2012}]{Bauswein2012}
{Bauswein} A.,  {Janka} H.-T.,  {Hebeler} K.,   {Schwenk} A.,  2012, \mn@doi
  [\prd] {10.1103/PhysRevD.86.063001}, \href
  {http://adsabs.harvard.edu/abs/2012PhRvD..86f3001B} {86, 063001}

\bibitem[\protect\citeauthoryear{{Bauswein}, {Goriely}  \& {Janka}}{{Bauswein}
  et~al.}{2013}]{Bauswein2013}
{Bauswein} A.,  {Goriely} S.,   {Janka} H.~T.,  2013, \mn@doi [\apj]
  {10.1088/0004-637X/773/1/78}, \href
  {https://ui.adsabs.harvard.edu/abs/2013ApJ...773...78B} {773, 78}

\bibitem[\protect\citeauthoryear{{Bernuzzi}, {Dietrich}  \& {Nagar}}{{Bernuzzi}
  et~al.}{2015}]{Bernuzzi2015}
{Bernuzzi} S.,  {Dietrich} T.,   {Nagar} A.,  2015, \mn@doi [\prl]
  {10.1103/PhysRevLett.115.091101}, \href
  {https://ui.adsabs.harvard.edu/abs/2015PhRvL.115i1101B} {115, 091101}

\bibitem[\protect\citeauthoryear{Bona, Masso, Seidel  \& Stela}{Bona
  et~al.}{1995}]{Bona:1994dr}
Bona C.,  Masso J.,  Seidel E.,   Stela J.,  1995, \mn@doi [Phys. Rev. Lett.]
  {10.1103/PhysRevLett.75.600}, 75, 600

\bibitem[\protect\citeauthoryear{Brown, Diener, Sarbach, Schnetter  \&
  Tiglio}{Brown et~al.}{2009}]{Brown:2008sb}
Brown J.~D.,  Diener P.,  Sarbach O.,  Schnetter E.,   Tiglio M.,  2009,
  \mn@doi [Phys. Rev. D] {10.1103/PhysRevD.79.044023}, 79, 044023

\bibitem[\protect\citeauthoryear{Carbone \& Schwenk}{Carbone \&
  Schwenk}{2019}]{Carbone:2019pkr}
Carbone A.,  Schwenk A.,  2019, \mn@doi [Phys. Rev. C]
  {10.1103/PhysRevC.100.025805}, 100, 025805

\bibitem[\protect\citeauthoryear{{Chatziioannou}}{{Chatziioannou}}{2020}]{Chatziioannou2020}
{Chatziioannou} K.,  2020, \mn@doi [General Relativity and Gravitation]
  {10.1007/s10714-020-02754-3}, \href
  {https://ui.adsabs.harvard.edu/abs/2020GReGr..52..109C} {52, 109}

\bibitem[\protect\citeauthoryear{{Constantinou}, {Muccioli}, {Prakash}  \&
  {Lattimer}}{{Constantinou} et~al.}{2015}]{Constantinou2015a}
{Constantinou} C.,  {Muccioli} B.,  {Prakash} M.,   {Lattimer} J.~M.,  2015,
  \mn@doi [\prc] {10.1103/PhysRevC.92.025801}, \href
  {http://adsabs.harvard.edu/abs/2015PhRvC..92b5801C} {92, 025801}

\bibitem[\protect\citeauthoryear{East, Paschalidis  \& Pretorius}{East
  et~al.}{2016a}]{East:2016zvv}
East W.~E.,  Paschalidis V.,   Pretorius F.,  2016a, \mn@doi [Class. Quant.
  Grav.] {10.1088/0264-9381/33/24/244004}, 33, 244004

\bibitem[\protect\citeauthoryear{East, Paschalidis, Pretorius  \& Shapiro}{East
  et~al.}{2016b}]{East:2015vix}
East W.~E.,  Paschalidis V.,  Pretorius F.,   Shapiro S.~L.,  2016b, \mn@doi
  [Phys. Rev. D] {10.1103/PhysRevD.93.024011}, 93, 024011

\bibitem[\protect\citeauthoryear{Espino, Bozzola  \& Paschalidis}{Espino
  et~al.}{2022}]{Espino2022}
Espino P.,  Bozzola G.,   Paschalidis V.,  2022, In preparation

\bibitem[\protect\citeauthoryear{{Etienne}, {Paschalidis}, {Haas}, {M{\"o}sta}
  \& {Shapiro}}{{Etienne} et~al.}{2015}]{Etienne2015}
{Etienne} Z.~B.,  {Paschalidis} V.,  {Haas} R.,  {M{\"o}sta} P.,   {Shapiro}
  S.~L.,  2015, \mn@doi [Classical and Quantum Gravity]
  {10.1088/0264-9381/32/17/175009}, \href
  {https://ui.adsabs.harvard.edu/abs/2015CQGra..32q5009E} {32, 175009}

\bibitem[\protect\citeauthoryear{{Figura}, {Lu}, {Burgio}, {Li}  \&
  {Schulze}}{{Figura} et~al.}{2020}]{Figura2020}
{Figura} A.,  {Lu} J.-J.,  {Burgio} G.~F.,  {Li} Z.-H.,   {Schulze} H.~J.,
  2020, \mn@doi [\prd] {10.1103/PhysRevD.102.043006}, \href
  {https://ui.adsabs.harvard.edu/abs/2020PhRvD.102d3006F} {102, 043006}

\bibitem[\protect\citeauthoryear{Foucart, O'Connor, Roberts, Kidder, Pfeiffer
  \& Scheel}{Foucart et~al.}{2016}]{Foucart:2016rxm}
Foucart F.,  O'Connor E.,  Roberts L.,  Kidder L.~E.,  Pfeiffer H.~P.,   Scheel
  M.~A.,  2016, \mn@doi [Phys. Rev. D] {10.1103/PhysRevD.94.123016}, 94, 123016

\bibitem[\protect\citeauthoryear{{Giacomazzo}, {Cipolletta}, {Kalinani},
  {Ciolfi}, {Sala}, {Giudici}  \& {Giangrandi}}{{Giacomazzo}
  et~al.}{2020}]{2020zndo...3689751G}
{Giacomazzo} B.,  {Cipolletta} F.,  {Kalinani} J.,  {Ciolfi} R.,  {Sala} L.,
  {Giudici} B.,   {Giangrandi} E.,  2020, {The Spritz Code},
  \mn@doi{10.5281/zenodo.3689751}

\bibitem[\protect\citeauthoryear{Hammond, Hawke  \& Andersson}{Hammond
  et~al.}{2021}]{Hammond:2021vtv}
Hammond P.,  Hawke I.,   Andersson N.,  2021, \mn@doi [Phys. Rev. D]
  {10.1103/PhysRevD.104.103006}, 104, 103006

\bibitem[\protect\citeauthoryear{{Hempel} \& {Schaffner-Bielich}}{{Hempel} \&
  {Schaffner-Bielich}}{2010}]{Hempel2010}
{Hempel} M.,  {Schaffner-Bielich} J.,  2010, \mn@doi [Nuclear Physics A]
  {10.1016/j.nuclphysa.2010.02.010}, \href
  {http://adsabs.harvard.edu/abs/2010NuPhA.837..210H} {837, 210}

\bibitem[\protect\citeauthoryear{{Hempel}, {Fischer}, {Schaffner-Bielich}  \&
  {Liebend{\"o}rfer}}{{Hempel} et~al.}{2012}]{Hempel2012}
{Hempel} M.,  {Fischer} T.,  {Schaffner-Bielich} J.,   {Liebend{\"o}rfer} M.,
  2012, \mn@doi [\apj] {10.1088/0004-637X/748/1/70}, \href
  {http://adsabs.harvard.edu/abs/2012ApJ...748...70H} {748, 70}

\bibitem[\protect\citeauthoryear{{Iosif} \& {Stergioulas}}{{Iosif} \&
  {Stergioulas}}{2021}]{2021MNRAS.503..850I}
{Iosif} P.,  {Stergioulas} N.,  2021, \mn@doi [\mnras] {10.1093/mnras/stab392},
  \href {https://ui.adsabs.harvard.edu/abs/2021MNRAS.503..850I} {503, 850}

\bibitem[\protect\citeauthoryear{{Janka}, {Zwerger}  \& {Moenchmeyer}}{{Janka}
  et~al.}{1993}]{Janka1993}
{Janka} H.-T.,  {Zwerger} T.,   {Moenchmeyer} R.,  1993, \aap, \href
  {http://adsabs.harvard.edu/abs/1993A%26A...268..360J} {268, 360}

\bibitem[\protect\citeauthoryear{Keller, Wellenhofer, Hebeler  \&
  Schwenk}{Keller et~al.}{2021}]{Keller:2020qhx}
Keller J.,  Wellenhofer C.,  Hebeler K.,   Schwenk A.,  2021, \mn@doi [Phys.
  Rev. C] {10.1103/PhysRevC.103.055806}, 103, 055806

\bibitem[\protect\citeauthoryear{{Lattimer} \& {Swesty}}{{Lattimer} \&
  {Swesty}}{1991}]{Lattimer1991}
{Lattimer} J.~M.,  {Swesty} D.~F.,  1991, \mn@doi [Nuclear Physics A]
  {10.1016/0375-9474(91)90452-C}, \href
  {http://adsabs.harvard.edu/abs/1991NuPhA.535..331L} {535, 331}

\bibitem[\protect\citeauthoryear{Lehner, Liebling, Palenzuela  \& Motl}{Lehner
  et~al.}{2016}]{Lehner:2016wjg}
Lehner L.,  Liebling S.~L.,  Palenzuela C.,   Motl P.~M.,  2016, \mn@doi [Phys.
  Rev. D] {10.1103/PhysRevD.94.043003}, 94, 043003

\bibitem[\protect\citeauthoryear{Loffler et~al.,}{Loffler
  et~al.}{2012}]{Loffler2012ETK}
Loffler F.,  et~al., 2012, \mn@doi [Classical and Quantum Gravity]
  {10.1088/0264-9381/29/11/115001}, 29, 115001

\bibitem[\protect\citeauthoryear{{Most} \& {Raithel}}{{Most} \&
  {Raithel}}{2021}]{Most2021}
{Most} E.~R.,  {Raithel} C.~A.,  2021, arXiv e-prints, \href
  {https://ui.adsabs.harvard.edu/abs/2021arXiv210706804M} {p. arXiv:2107.06804}

\bibitem[\protect\citeauthoryear{{Most}, {Papenfort}  \& {Rezzolla}}{{Most}
  et~al.}{2019}]{Most2019}
{Most} E.~R.,  {Papenfort} L.~J.,   {Rezzolla} L.,  2019, \mn@doi [\mnras]
  {10.1093/mnras/stz2809}, \href
  {https://ui.adsabs.harvard.edu/abs/2019MNRAS.490.3588M} {490, 3588}

\bibitem[\protect\citeauthoryear{{Most} et~al.,}{{Most}
  et~al.}{2022}]{Most2022}
{Most} E.~R.,  et~al., 2022, \mn@doi [\mnras] {10.1093/mnras/stab2793}, \href
  {https://ui.adsabs.harvard.edu/abs/2022MNRAS.509.1096M} {509, 1096}

\bibitem[\protect\citeauthoryear{{Mosta} et~al.,}{{Mosta}
  et~al.}{2014}]{2014CQGra..31a5005M}
{Mosta} P.,  et~al., 2014, \mn@doi [Classical and Quantum Gravity]
  {10.1088/0264-9381/31/1/015005}, \href
  {https://ui.adsabs.harvard.edu/abs/2014CQGra..31a5005M} {31, 015005}

\bibitem[\protect\citeauthoryear{Nakamura, Oohara  \& Kojima}{Nakamura
  et~al.}{1987}]{BSSN1}
Nakamura T.,  Oohara K.,   Kojima Y.,  1987, \mn@doi [Prog. Theor. Phys.
  Suppl.] {10.1143/PTPS.90.1}, 90, 1

\bibitem[\protect\citeauthoryear{Newman \& Penrose}{Newman \&
  Penrose}{1962}]{Newman:1961qr}
Newman E.,  Penrose R.,  1962, \mn@doi [J. Math. Phys.] {10.1063/1.1724257}, 3,
  566

\bibitem[\protect\citeauthoryear{{Oechslin}, {Janka}  \& {Marek}}{{Oechslin}
  et~al.}{2007}]{Oechslin2007}
{Oechslin} R.,  {Janka} H.-T.,   {Marek} A.,  2007, \mn@doi [\aap]
  {10.1051/0004-6361:20066682}, \href
  {http://adsabs.harvard.edu/abs/2007A%26A...467..395O} {467, 395}

\bibitem[\protect\citeauthoryear{{Oertel}, {Hempel}, {Kl{\"a}hn}  \&
  {Typel}}{{Oertel} et~al.}{2017}]{Oertel2017}
{Oertel} M.,  {Hempel} M.,  {Kl{\"a}hn} T.,   {Typel} S.,  2017, \mn@doi
  [Reviews of Modern Physics] {10.1103/RevModPhys.89.015007}, \href
  {http://adsabs.harvard.edu/abs/2017RvMP...89a5007O} {89, 015007}

\bibitem[\protect\citeauthoryear{{Oppenheimer} \& {Volkoff}}{{Oppenheimer} \&
  {Volkoff}}{1939}]{Oppenheimer1939}
{Oppenheimer} J.~R.,  {Volkoff} G.~M.,  1939, \mn@doi [Physical Review]
  {10.1103/PhysRev.55.374}, \href
  {http://adsabs.harvard.edu/abs/1939PhRv...55..374O} {55, 374}

\bibitem[\protect\citeauthoryear{{{\"O}zel} \& {Freire}}{{{\"O}zel} \&
  {Freire}}{2016}]{Ozel2016}
{{\"O}zel} F.,  {Freire} P.,  2016, \mn@doi [\araa]
  {10.1146/annurev-astro-081915-023322}, \href
  {http://adsabs.harvard.edu/abs/2016ARA%26A..54..401O} {54, 401}

\bibitem[\protect\citeauthoryear{{{\"O}zel} \& {Psaltis}}{{{\"O}zel} \&
  {Psaltis}}{2009}]{Ozel2009}
{{\"O}zel} F.,  {Psaltis} D.,  2009, \mn@doi [\prd]
  {10.1103/PhysRevD.80.103003}, \href
  {http://adsabs.harvard.edu/abs/2009PhRvD..80j3003O} {80, 103003}

\bibitem[\protect\citeauthoryear{Palenzuela, Liebling, Neilsen, Lehner,
  Caballero, O?Connor  \& Anderson}{Palenzuela et~al.}{2015}]{Palenzuela_2015}
Palenzuela C.,  Liebling S.~L.,  Neilsen D.,  Lehner L.,  Caballero O.,
  O?Connor E.,   Anderson M.,  2015, \mn@doi [Physical Review D]
  {10.1103/physrevd.92.044045}, 92

\bibitem[\protect\citeauthoryear{{Paschalidis} \& {Stergioulas}}{{Paschalidis}
  \& {Stergioulas}}{2017}]{Paschalidis2017}
{Paschalidis} V.,  {Stergioulas} N.,  2017, \mn@doi [Living Reviews in
  Relativity] {10.1007/s41114-017-0008-x}, \href
  {http://adsabs.harvard.edu/abs/2017LRR....20....7P} {20, 7}

\bibitem[\protect\citeauthoryear{{Paschalidis}, {Etienne}  \&
  {Shapiro}}{{Paschalidis} et~al.}{2012}]{Paschalidis2012}
{Paschalidis} V.,  {Etienne} Z.~B.,   {Shapiro} S.~L.,  2012, \mn@doi [\prd]
  {10.1103/PhysRevD.86.064032}, \href
  {http://adsabs.harvard.edu/abs/2012PhRvD..86f4032P} {86, 064032}

\bibitem[\protect\citeauthoryear{Paschalidis, East, Pretorius  \&
  Shapiro}{Paschalidis et~al.}{2015}]{Paschalidis:2015mla}
Paschalidis V.,  East W.~E.,  Pretorius F.,   Shapiro S.~L.,  2015, \mn@doi
  [Phys. Rev. D] {10.1103/PhysRevD.92.121502}, 92, 121502

\bibitem[\protect\citeauthoryear{Penrose}{Penrose}{1963}]{Penrose:1962ij}
Penrose R.,  1963, \mn@doi [Phys. Rev. Lett.] {10.1103/PhysRevLett.10.66}, 10,
  66

\bibitem[\protect\citeauthoryear{Radice, Rezzolla  \& Galeazzi}{Radice
  et~al.}{2014}]{Radice_2014}
Radice D.,  Rezzolla L.,   Galeazzi F.,  2014, \mn@doi [Classical and Quantum
  Gravity] {10.1088/0264-9381/31/7/075012}, 31, 075012

\bibitem[\protect\citeauthoryear{Radice, Bernuzzi  \& Ott}{Radice
  et~al.}{2016a}]{Radice:2016gym}
Radice D.,  Bernuzzi S.,   Ott C.~D.,  2016a, \mn@doi [Phys. Rev. D]
  {10.1103/PhysRevD.94.064011}, 94, 064011

\bibitem[\protect\citeauthoryear{Radice, Galeazzi, Lippuner, Roberts, Ott  \&
  Rezzolla}{Radice et~al.}{2016b}]{Radice:2016dwd}
Radice D.,  Galeazzi F.,  Lippuner J.,  Roberts L.~F.,  Ott C.~D.,   Rezzolla
  L.,  2016b, \mn@doi [Mon. Not. Roy. Astron. Soc.] {10.1093/mnras/stw1227},
  460, 3255

\bibitem[\protect\citeauthoryear{Radice, Bernuzzi, Perego  \& Haas}{Radice
  et~al.}{2022}]{Radice:2021jtw}
Radice D.,  Bernuzzi S.,  Perego A.,   Haas R.,  2022, \mn@doi [Mon. Not. Roy.
  Astron. Soc.] {10.1093/mnras/stac589}, 512, 1499

\bibitem[\protect\citeauthoryear{{Raithel}}{{Raithel}}{2019}]{Raithel2019a}
{Raithel} C.~A.,  2019, \mn@doi [European Physical Journal A]
  {10.1140/epja/i2019-12759-5}, \href
  {https://ui.adsabs.harvard.edu/abs/2019EPJA...55...80R} {55, 80}

\bibitem[\protect\citeauthoryear{{Raithel} \& {Most}}{{Raithel} \&
  {Most}}{2022}]{Raithel2022}
{Raithel} C.~A.,  {Most} E.~R.,  2022, arXiv e-prints, \href
  {https://ui.adsabs.harvard.edu/abs/2022arXiv220103594R} {p. arXiv:2201.03594}

\bibitem[\protect\citeauthoryear{Raithel \& Paschalidis}{Raithel \&
  Paschalidis}{2022}]{Raithel:2022san}
Raithel C.~A.,  Paschalidis V.,  2022

\bibitem[\protect\citeauthoryear{{Raithel}, {{\"O}zel}  \& {Psaltis}}{{Raithel}
  et~al.}{2019}]{Raithel2019}
{Raithel} C.~A.,  {{\"O}zel} F.,   {Psaltis} D.,  2019, \mn@doi [\apj]
  {10.3847/1538-4357/ab08ea}, \href
  {https://ui.adsabs.harvard.edu/abs/2019ApJ...875...12R} {875, 12}

\bibitem[\protect\citeauthoryear{{Raithel}, {Paschalidis}  \&
  {{\~A}-zel}}{{Raithel} et~al.}{2021a}]{Raithel2021}
{Raithel} C.~A.,  {Paschalidis} V.,   {{\~A}-zel} F.,  2021a, \mn@doi [\prd]
  {10.1103/PhysRevD.104.063016}, \href
  {https://ui.adsabs.harvard.edu/abs/2021PhRvD.104f3016R} {104, 063016}

\bibitem[\protect\citeauthoryear{{Raithel}, {{\"O}zel}  \& {Psaltis}}{{Raithel}
  et~al.}{2021b}]{ROPerratum}
{Raithel} C.~A.,  {{\"O}zel} F.,   {Psaltis} D.,  2021b, \mn@doi [\apj]
  {10.3847/1538-4357/ac0630}, \href
  {https://ui.adsabs.harvard.edu/abs/2021ApJ...915...73R} {915, 73}

\bibitem[\protect\citeauthoryear{{Read}, {Lackey}, {Owen}  \&
  {Friedman}}{{Read} et~al.}{2009}]{Read2009}
{Read} J.~S.,  {Lackey} B.~D.,  {Owen} B.~J.,   {Friedman} J.~L.,  2009,
  \mn@doi [\prd] {10.1103/PhysRevD.79.124032}, \href
  {http://adsabs.harvard.edu/abs/2009PhRvD..79l4032R} {79, 124032}

\bibitem[\protect\citeauthoryear{{Reisswig} \& {Pollney}}{{Reisswig} \&
  {Pollney}}{2011}]{Reisswig2011}
{Reisswig} C.,  {Pollney} D.,  2011, \mn@doi [Classical and Quantum Gravity]
  {10.1088/0264-9381/28/19/195015}, \href
  {https://ui.adsabs.harvard.edu/abs/2011CQGra..28s5015R} {28, 195015}

\bibitem[\protect\citeauthoryear{Reisswig, Ott, Sperhake  \&
  Schnetter}{Reisswig et~al.}{2011}]{Reisswig_2011GWs}
Reisswig C.,  Ott C.~D.,  Sperhake U.,   Schnetter E.,  2011, \mn@doi [Physical
  Review D] {10.1103/physrevd.83.064008}, 83

\bibitem[\protect\citeauthoryear{Rosswog \& Liebendoerfer}{Rosswog \&
  Liebendoerfer}{2003}]{Rosswog:2003rv}
Rosswog S.,  Liebendoerfer M.,  2003, \mn@doi [Mon. Not. Roy. Astron. Soc.]
  {10.1046/j.1365-8711.2003.06579.x}, 342, 673

\bibitem[\protect\citeauthoryear{{Schneider}, {Roberts}  \& {Ott}}{{Schneider}
  et~al.}{2017}]{Schneider2017}
{Schneider} A.~S.,  {Roberts} L.~F.,   {Ott} C.~D.,  2017, \mn@doi [\prc]
  {10.1103/PhysRevC.96.065802}, \href
  {http://adsabs.harvard.edu/abs/2017PhRvC..96f5802S} {96, 065802}

\bibitem[\protect\citeauthoryear{{Sekiguchi}, {Kiuchi}, {Kyutoku}  \&
  {Shibata}}{{Sekiguchi} et~al.}{2011}]{Sekiguchi2011}
{Sekiguchi} Y.,  {Kiuchi} K.,  {Kyutoku} K.,   {Shibata} M.,  2011, \mn@doi
  [Physical Review Letters] {10.1103/PhysRevLett.107.051102}, \href
  {http://adsabs.harvard.edu/abs/2011PhRvL.107e1102S} {107, 051102}

\bibitem[\protect\citeauthoryear{{Shen}, {Toki}, {Oyamatsu}  \&
  {Sumiyoshi}}{{Shen} et~al.}{1998}]{Shen1998a}
{Shen} H.,  {Toki} H.,  {Oyamatsu} K.,   {Sumiyoshi} K.,  1998, \mn@doi
  [Nuclear Physics A] {10.1016/S0375-9474(98)00236-X}, \href
  {http://adsabs.harvard.edu/abs/1998NuPhA.637..435S} {637, 435}

\bibitem[\protect\citeauthoryear{{Shibata} \& {Nakamura}}{{Shibata} \&
  {Nakamura}}{1995}]{Shibata1995}
{Shibata} M.,  {Nakamura} T.,  1995, \mn@doi [\prd] {10.1103/PhysRevD.52.5428},
  \href {https://ui.adsabs.harvard.edu/abs/1995PhRvD..52.5428S} {52, 5428}

\bibitem[\protect\citeauthoryear{Siegel, Mosta, Desai  \& Wu}{Siegel
  et~al.}{2018}]{Siegel:2017sav}
Siegel D.~M.,  Mosta P.,  Desai D.,   Wu S.,  2018, \mn@doi [Astrophys. J.]
  {10.3847/1538-4357/aabcc5}, 859, 71

\bibitem[\protect\citeauthoryear{{Steiner}, {Lattimer}  \& {Brown}}{{Steiner}
  et~al.}{2013}]{Steiner2013}
{Steiner} A.~W.,  {Lattimer} J.~M.,   {Brown} E.~F.,  2013, \mn@doi [\apjl]
  {10.1088/2041-8205/765/1/L5}, \href
  {http://adsabs.harvard.edu/abs/2013ApJ...765L...5S} {765, L5}

\bibitem[\protect\citeauthoryear{{Takami}, {Rezzolla}  \& {Baiotti}}{{Takami}
  et~al.}{2014}]{Takami2014}
{Takami} K.,  {Rezzolla} L.,   {Baiotti} L.,  2014, \mn@doi [Physical Review
  Letters] {10.1103/PhysRevLett.113.091104}, \href
  {http://adsabs.harvard.edu/abs/2014PhRvL.113i1104T} {113, 091104}

\bibitem[\protect\citeauthoryear{{Tolman}}{{Tolman}}{1939}]{Tolman1939}
{Tolman} R.~C.,  1939, \mn@doi [Physical Review] {10.1103/PhysRev.55.364},
  \href {http://adsabs.harvard.edu/abs/1939PhRv...55..364T} {55, 364}

\bibitem[\protect\citeauthoryear{Tonetto \& Benhar}{Tonetto \&
  Benhar}{2022}]{Tonetto:2022zhs}
Tonetto L.,  Benhar O.,  2022

\bibitem[\protect\citeauthoryear{{Typel}, {Oertel}  \& {Kl{\"a}hn}}{{Typel}
  et~al.}{2015}]{Typel2015}
{Typel} S.,  {Oertel} M.,   {Kl{\"a}hn} T.,  2015, \mn@doi [Physics of
  Particles and Nuclei] {10.1134/S1063779615040061}, \href
  {https://ui.adsabs.harvard.edu/abs/2015PPN....46..633T} {46, 633}

\bibitem[\protect\citeauthoryear{{Vretinaris}, {Stergioulas}  \&
  {Bauswein}}{{Vretinaris} et~al.}{2020}]{Vretinaris2020}
{Vretinaris} S.,  {Stergioulas} N.,   {Bauswein} A.,  2020, \mn@doi [\prd]
  {10.1103/PhysRevD.101.084039}, \href
  {https://ui.adsabs.harvard.edu/abs/2020PhRvD.101h4039V} {101, 084039}

\makeatother
\end{thebibliography}
\bibliographystyle{mnras}

\appendix

\section{Comparison of pressure approximations}
\label{sec:EoSappendix}
In this appendix, we present a brief comparison of the pressure approximation for SFHo.
A more detailed exploration of the errors introduced by the
 $M^*$-framework at finite-temperatures and arbitrary electrons
can be found in \ROP, for a larger sample of EoSs which includes SFHo.

\begin{figure*}
\centering
\includegraphics[width=0.8\textwidth]{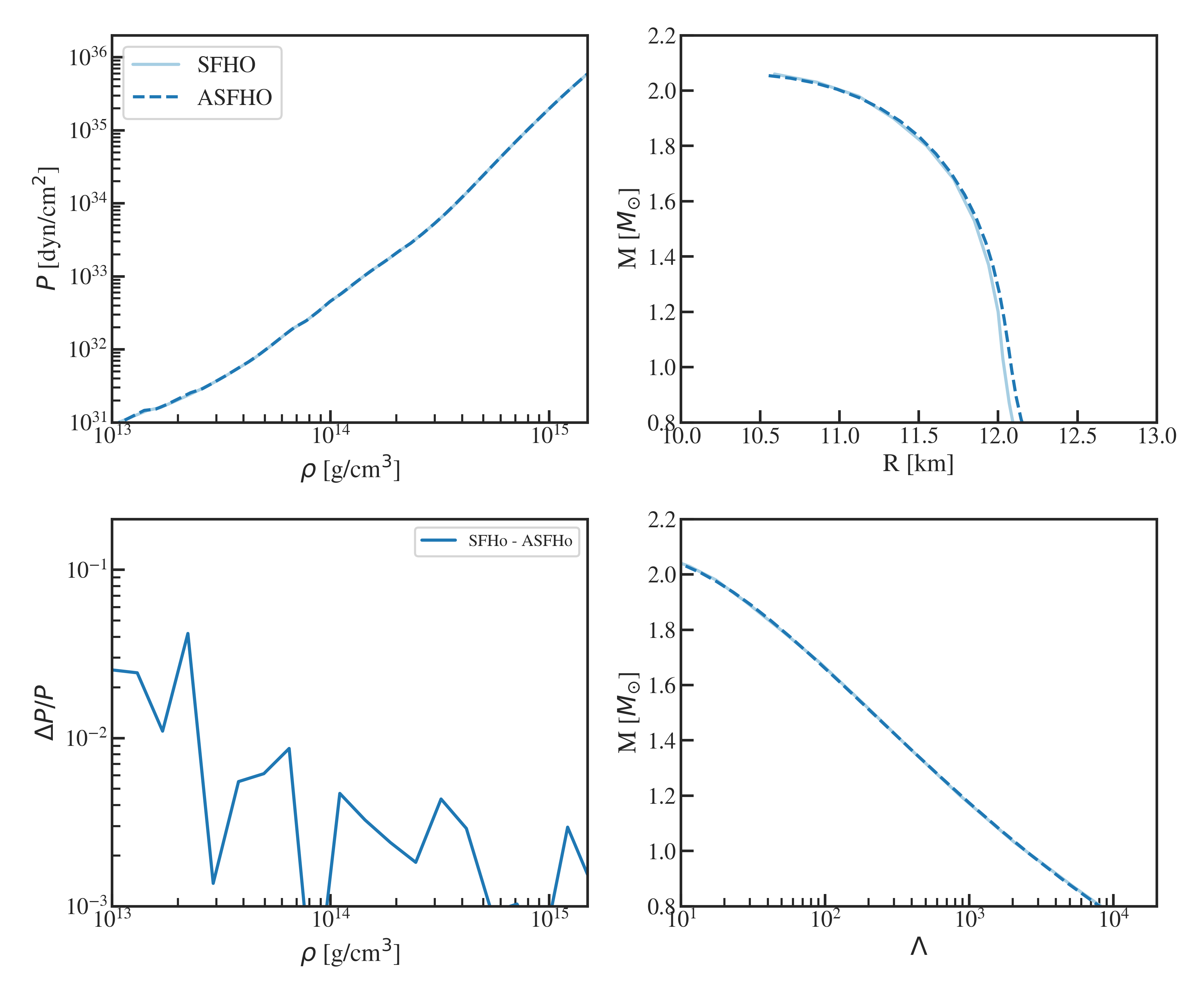}
\caption{\label{fig:EoS} Clockwise from top left: the cold, \beq~pressure for the full 
and approximate versions of SFHo;  the corresponding mass-radius relations; 
the tidal deformability;  and the fractional differences in 
pressure as a function of density. The solid, light-blue 
lines correspond to the full version of the SFHo (obtained from
 \texttt{stellarcollapse.org}), at $T=0.1$~MeV and in \beq.
 The dashed, dark-blue lines correspond to the approximate
 version of SFHo under the same conditions.}
\end{figure*}

 Figure~\ref{fig:EoS} shows the cold, \beq~ slices of the approximate 
 and full versions of SFHo, along with the corresponding
 mass-radius and mass-tidal deformability curves. Although the approximate
 EoS table is constructed starting from the cold, \beq~ slice of the full EoS
 (see Fig.~\ref{fig:construction}), the extrapolations to construct the 3D approximate table
 introduce some errors. As a result of small errors in the approximate
 chemical potentials, when we 
 extract the \beq~ slice from the 3D approximate table,
 it does not exactly match the starting slice. This can be seen in 
 the slight disagreement between the cold, \beq~ slices of 
 the EoSs shown in Fig.~\ref{fig:EoS}. 
 The error in the \beq~ slice is typically $\lesssim$1\%, at densities above $0.1\rns$.
 Slightly larger errors (up to a few percent) can be found at lower densities, 
 where the approximate EoS is being matched to the full table.  
 The impact of these errors on the global properties of the neutron star (i.e.,
 mass and radius) can be seen in the right panels of Fig.~\ref{fig:EoS}. 
 The radius of a 1.4~$\Ms$ neutron star differs by 0.03~km between
  the approximate and full EoS models (fractional difference of 0.3\%),
 while the corresponding tidal deformabilities differ by only $\sim0.02\%$. 
 Thus, the impact of the approximations on the properties of the cold, 
 equilibrium stars is negligible. 
  
\begin{figure*}
\centering
\includegraphics[width=0.8\textwidth]{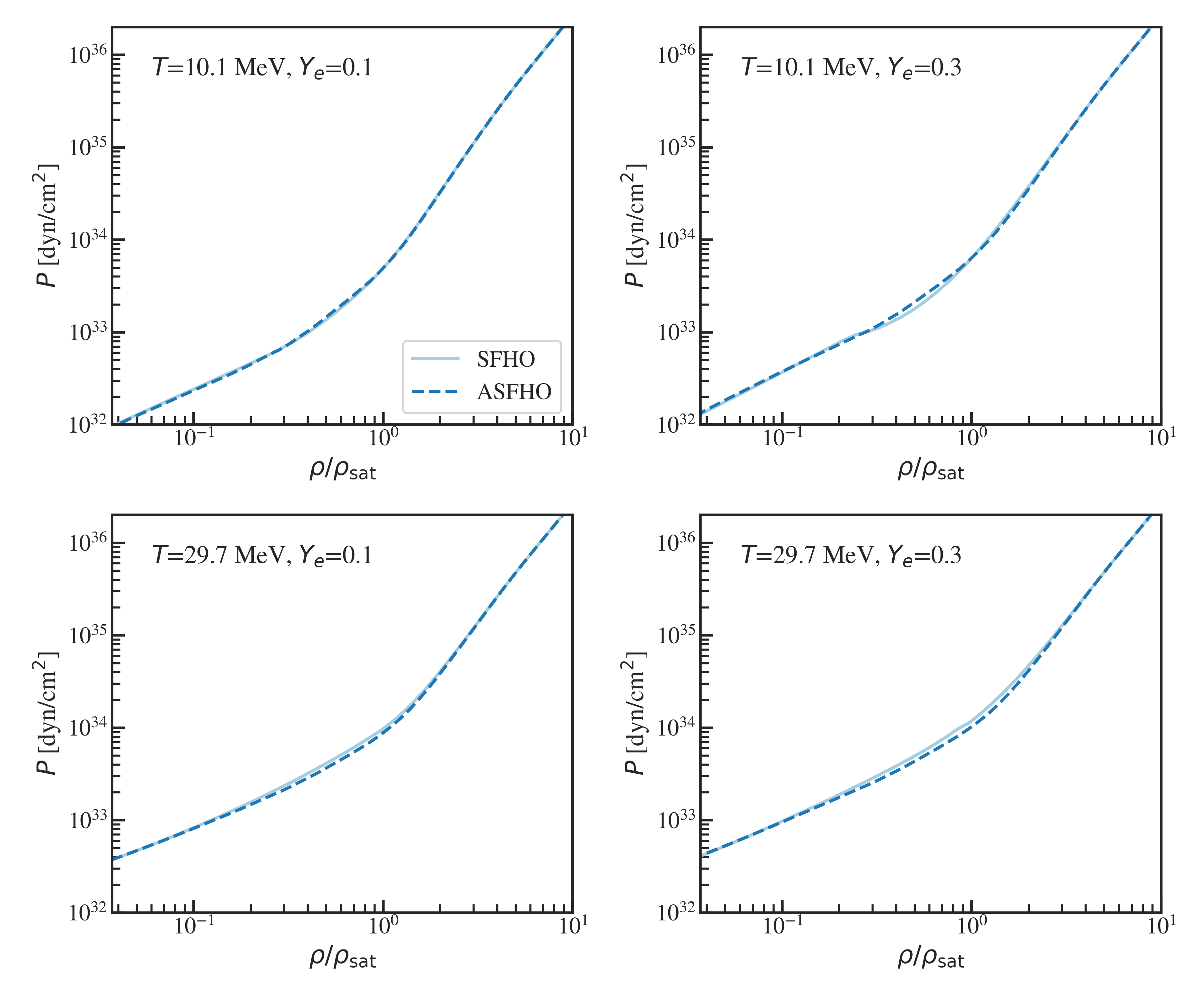}
\caption{\label{fig:Pmodel} Comparison of the pressure model at finite temperatures and fixed electron fractions.
The full version of SFHo is shown in light blue, while the approximate version
is shown with the dark dashed line. The approximate EoS starts from same cold, \beq~EoS and is extended
to finite temperature and a fixed (non-equilibrium) electron fraction using the $M^*$-framework.}
\end{figure*}

 Figure~\ref{fig:Pmodel} shows the EoS at two finite-temperatures and fixed, non-equilibrium 
 electron fractions. We choose temperatures of $T\simeq 10$ and 30~MeV, which are 
 motivated by the temperatures reached in the interior of the remnant at the merger. 
 The electron fractions are fixed to 0.1 or 0.3. We note that, because $Y_e^{\beta}$ is a 
 density-dependent function, fixing $Y_e$ to a constant value is a stringent test of the model,
 as this requires deviations from equilibrium at essentially all densities. Indeed, the errors
 of the approximate model are largest in this figure for $Y_e=0.3$ just below $\rho_{\rm sat}$. 
 At these densities, the electron fraction of the \beq~EoS is $\sim0.02$; thus, a value of $Y_e=0.3$
 is very far out of equilibrium, and the errors of the approximation are accordingly larger.
 Such extreme out-of-equilibrium conditions are not reached in our simulations.
 
 Additional comparisons between the full and approximate EoSs for a larger sample of 
 models can be found in \ROP. We recreate the SFHo result here, to illustrate the 
 general agreement between the approximate 
 and full EoSs across a range of densities and temperatures of interest in a merger.

\end{document}